\documentclass[twocolumn,showpacs,preprintnumbers,amsmath,amssymb, prx, reprint, longbibliography]{revtex4-1}
\usepackage{amsmath}
\usepackage{stmaryrd}
\usepackage{txfonts}
\usepackage{amssymb}
\usepackage{mathrsfs}
\usepackage{graphicx}
\usepackage{dcolumn}
\usepackage{bm}
\usepackage{epsfig}
\usepackage{color}
\usepackage{pbox}
\usepackage[colorlinks=true, linkcolor=blue, citecolor=blue, urlcolor=blue]{hyperref} 
\begin{document}

\title{Ginzburg-Landau theory for skyrmions in inversion-symmetric magnets with competing interactions}
\author{Shi-Zeng Lin}
\email{szl@lanl.gov}
\affiliation{Theoretical Division, Los Alamos National Laboratory, Los Alamos, New Mexico 87545, USA}
\author{Satoru Hayami}
\affiliation{Theory Division, T-4 and CNLS, Los Alamos National Laboratory, Los Alamos, NM 87545, USA}

\begin{abstract}
Magnetic skyrmions have attracted considerable attention recently for their huge potential in spintronic applications. Generally skyrmions are big compared to the atomic lattice constant, which allows for the Ginzburg-Landau type description in the continuum limit. Such a description successfully captures the main experimental observations on skyrmions in B20 compound without inversion symmetry. Skyrmions can also exist in inversion-symmetric magnets with competing interactions. Here we derive a general Ginzburg-Landau theory for skyrmions in these magnets valid in the long wavelength limit. We study the unusual static and dynamical properties of skyrmions based on the derived Ginzburg-Landau theory. We show that an easy axis spin anisotropy is sufficient to stabilize a skyrmion lattice. Interestingly, the skyrmion in inversion-symmetric magnets has a new internal degree of freedom associated with the rotation of helicity, i.e. the ``spin" of the skyrmion as a particle, in addition to the usual translational motion of skyrmions (orbital motion). The orbital and spin degree of freedoms of an individual skyrmion can couple to each other, and give rise to unusual behavior that is absent for the skyrmions stabilized by the Dzyaloshinskii-Moriya interaction. The derived Ginzburg-Landau theory provides a convenient and general framework to discuss skyrmion physics and will facilitate the search for skyrmions in inversion-symmetric magnets.
\end{abstract}
\date{\today}
\maketitle

\section{Introduction}

A magnetic skyrmion is a topologically protected excitation of spin texture in magnets \cite{Bogdanov89,Bogdanov94}. Skyrmions were discovered in B20 compound without inversion symmetry in 2009 \cite{Muhlbauer2009}. Due to their unique topology, skyrmions give birth to many emergent phenomena, such as topological Hall effect  \cite{PhysRevLett.110.117202,Neubauer2009}, magnetoelectric coupling \cite{Seki2012,Seki2012b} etc. A skyrmion behaving as a particle can also be manipulated in a controlled way by electric current \cite{Jonietz2010,Yu2012,Schulz2012}, thermal gradient \cite{Kong2013,Lin2014PRL,Mochizuki2014}, strain \cite{Shibata2015} and so on. Especially the threshold current to drive skyrmion into motion is 5 or 6 orders of magnitude lower than that of magnetic domain walls \cite{Jonietz2010,Yu2012,Schulz2012}. For these unique properties, skyrmions are deemed as a prime candidate for applications in next-generation spintronic devices and therefore have attracted tremendous attention recently. \cite{Fert2013,nagaosa_topological_2013}

Experiments with different imaging methods have firmly established skyrmions as a ubiquitous state in magnets. Skyrmions were found in metals \cite{Muhlbauer2009,Yu2010a}, semiconductors \cite{Yu2011} and insulators \cite{Adams2012,Seki2012}. More strikingly, the phase diagram for bulks or thin films in these materials are similar, implying that the magnetic properties are governed by the same low energy effective theory because the size of a skyrmion is much bigger than the atomic lattice constant. Such a phenomenological Ginzburg-Landau theory was proposed by Bak and Jensen long time ago based on symmetry consideration and expansion in the ordering wave vector \cite{bak_theory_1980}. Experiments and theoretical calculations have demonstrated that the Ginzburg-Landau theory correctly captures the main features of the skyrmion physics in B20 compounds \cite{Yu2010a,Mochizuki2012,PhysRevLett.109.037603,Mochizuki2013,okamura_microwave_2013,schwarze_universal_2015,Zhang2015}. 
    
The majority of the skyrmion hosting materials discussed so far do not have inversion symmetry, where the Dzyaloshinskii-Moriya (DM) interaction \cite{Dzyaloshinsky1958,Moriya60,Moriya60b} stabilizes the skyrmion phase. To render a skyrmion (meta)-stable, one necessary condition is to introduce a characteristic length scale in the system in accordance with the Derrick's theorem \cite{Derrick1252}, thus requires competing interactions. The DM interaction is one way to introduce a length scale. In fact, it was found experimentally that the skyrmion lattice can also be stabilized by a long-range dipolar interaction. \cite{Yu2012b} The competing interaction in frustrated magnets with inversion symmetry is another route to stabilize skyrmions.

A triangular lattice of skyrmions can be regarded as a superposition of three helices with the ordering wave vector rotated by $120^{\circ}$. The Heisenberg model with competing interactions on a triangular lattice then becomes an ideal system to realize the skyrmion lattice. First the competing interactions can produce a magnetic helix. Meanwhile the triangular lattice itself provides a spatial anisotropy to align the helix along the three equivalent directions. At low temperatures, the superposition of three helices violates the constraint that the total moment $|S|$ is constant in space, thus costs energy. However at high temperatures, the moment becomes soft and the skyrmion lattice is favored. Indeed the skyrmion lattice was found in the frustrated Heisenberg model on a triangular lattice at a nonzero temperature in Ref. \onlinecite{PhysRevLett.108.017206}. However it was unclear whether a single skyrmion can be stabilized or not.

Recently, Leonov and Mostovoy studied the skyrmions in the frustrated Heisenberg model on a triangular lattice with an easy axis anisotropy \cite{leonov_multiply_2015}. They found that an easy axis anisotropy is helpful to stabilize the skyrmion lattice, similar to that with DM interaction \cite{Butenko2010,Wilson2012,Huang2012}. They also found the existence of an isolated skyrmion. The helicity of skyrmion oscillates when one moves away from the the skyrmion center. The interaction between skyrmions is nonmonotonic as a function of distance and depends on the helicity. Moreover there exists a collective mode associated with the helicity of the skyrmion, which can couple to the skyrmion center of mass motion and gives rise to a dynamical magnetoelectric effect.

Generally the skyrmion size is much bigger than the lattice parameter of the spin system. The spin lattice should not matter because skyrmions decouple from the spin lattice. Such a large skyrmion limit also allows for a universal description in the continuum limit similar to the systems with the DM interaction. In this work, we present a Ginzburg-Landau theory for skyrmions in inversion symmetric magnets (ISM) with competing interactions. The Ginzburg-Landau energy is obtained by expansion in term of the ordering wave vector which is small compared to $2\pi/a$, with $a$ being the lattice parameter of the spin system. The Ginzburg-Landau approach enables us to describe the properties of skyrmions in a more transparent way. The Ginzburg-Landau theory reproduces some of the results already discussed in Ref. \onlinecite{leonov_multiply_2015}. Most importantly, we demonstrate explicitly that it is sufficient to stabilize a skyrmion lattice with an easy axis spin anisotropy and competing interactions between spins for any spin lattice. The skyrmion in ISM has a new internal degree of freedom associated with the global rotation of spin along the magnetic field axis, or the ``spin" of the skyrmion. This spin degree of freedom can couple with the orbital degree of freedom (translational motion) of skyrmions. Specifically, first we find that the size of skyrmion diverges when the field approaches the saturation field. Secondly, we find that the rotation of helicity is coupled with the translation motion of skyrmion. In the presence of a spin Hall torque, the skyrmion moves along a circle. Thirdly we present unbiased Monte Carlo simulations and find that the skyrmion lattice is stable in certain region in the magnetic field-temperature phase diagram in the presence of an easy axis anisotropy. Finally we demonstrate that skyrmions and antiskyrmions can be created by annealing. The Ginzburg-Landau description provides a theoretical basis for understanding skyrmions in ISM with competing interactions. Meanwhile, we demonstrate novel properties of skyrmions in these magnets, which are not shared by those in systems DM interaction.

The remaining of the paper is organized as follows. In Sec. \ref{Sec2} we present the Ginzburg-Landau theory. In Sec. \ref{Sec3}, we discuss a single skyrmion excitation in the ferromagnetic background. In Sec. \ref{Sec4}. we study the pairwise interaction between skyrmions and/or antiskyrmions. We then show in Sec. \ref{Sec5} that skyrmions and antiskyrmions can be excited by annealing. In Sec. \ref{Sec6} we investigate the thermodynamically stable skyrmion lattice phase favored by an easy axis anisotropy. In Sec. \ref{Sec7} we consider the dynamics of a single skyrmion driven by a torque generated by an electric current. The paper is concluded by brief discussions and summary in Sec. \ref{Sec8}.

\section{Ginzburg-Landau theory}\label{Sec2}

We consider the inversion-symmetric classical Heisenberg models with competing interactions, where the competing interaction can stabilize a state with nonzero ordering wave vector $Q$ as a ground state. We focus on the limit when $Q a <<1$, which allows us to take the continuum limit. Then the skyrmion size is much bigger than the lattice parameter of the spin system, and the skyrmions are  decoupled from the underlying spin lattice. Expanding the Hamiltonian to the quartic order in $Q$, we obtain the Ginzburg-Landau energy
\begin{equation}\label{eqs1}
\mathcal{H}=\int dr^3\left[-\frac{I_1}{2} (\nabla \mathbf{S})^2+\frac{I_2}{2}(\nabla^2 \mathbf{S})^2-\mathbf{H}_a\cdot\mathbf{S}\right],
\end{equation}
with the constraint that $|\mathbf{S}|=1$. The last term represents the Zeeman interaction with an external magnetic field $\mathbf{H}_a$. For systems with spatial inversion symmetry, only the terms with even power in $Q$ is allowed. When $I_1<0$, we have $Q=0$ and we can neglect the $(\nabla^2 \mathbf{S})^2$ term, therefore Eq. \eqref{eqs1} reduces to the standard nonlinear $\sigma$ model. We focus on the interesting regime when $I_1>0$, where the competing interactions introduce a length scale and stabilize the skyrmion solution as will be discussed below. We introduce dimensionless units by normalizing length in unit of $\sqrt{I_2/I_1}$ and energy density in unit of $I_1^2/I_2$. Then Eq. \eqref{eqs1} can be casted into
\begin{equation}\label{eqs2}
\mathcal{H}=\int dr^3\left[-\frac{1}{2} (\nabla \mathbf{S})^2+\frac{1}{2}(\nabla^2 \mathbf{S})^2-\mathbf{H}_a\cdot\mathbf{S}\right].
\end{equation}
The competing interactions yield a characteristic length scale $Q_0=1/\sqrt{2}$, which is a necessary condition for stabilizing the skyrmion solution according to the Derrick's theorem \cite{Derrick1252}. The direction of $Q_0$ is not determined and can be fixed by higher order terms neglected here, such as spatial anisotropy. The Hamiltonian Eq. \eqref{eqs2} is invariant under the following operations: inversion, translation and global rotation of spin along the direction of magnetic field. The inversion symmetry indicates that a skyrmion and an antiskyrmion with opposite winding direction are degenerate in energy. The translational invariant and $U(1)$ symmetry mean that there are two Goldstone modes associated with the translational motion and rotation of the skyrmion as will be discussed in Sec. \ref{Sec7} in details.

At zero temperature, the ground state spin configuration is a conical spiral with $\mathbf{S}=[\sin\theta\cos(\mathbf{Q}_0\cdot \mathbf{r}),\ \sin\theta\sin(\mathbf{Q}_0\cdot \mathbf{r}),\ \cos\theta]$ when $H_a\le H_s$ with $H_s=1/4$ being the saturation field. The canting angle $\theta$ is given by $\cos\theta=H_a/H_s$. When $H_a\ge H_s$, the system becomes a fully polarized ferromagnetic (FM) state via a second order phase transition. No thermodynamically stable skyrmion lattice is allowed according to the Hamiltonian Eq. \eqref{eqs2}. A skyrmion can exist as a metastable state (see Sec. \ref{Sec3}) or a skyrmion lattice can be stabilized by an easy axis spin anisotropy (see Sec. \ref{Sec6}).

\begin{figure}[t]
\psfig{figure=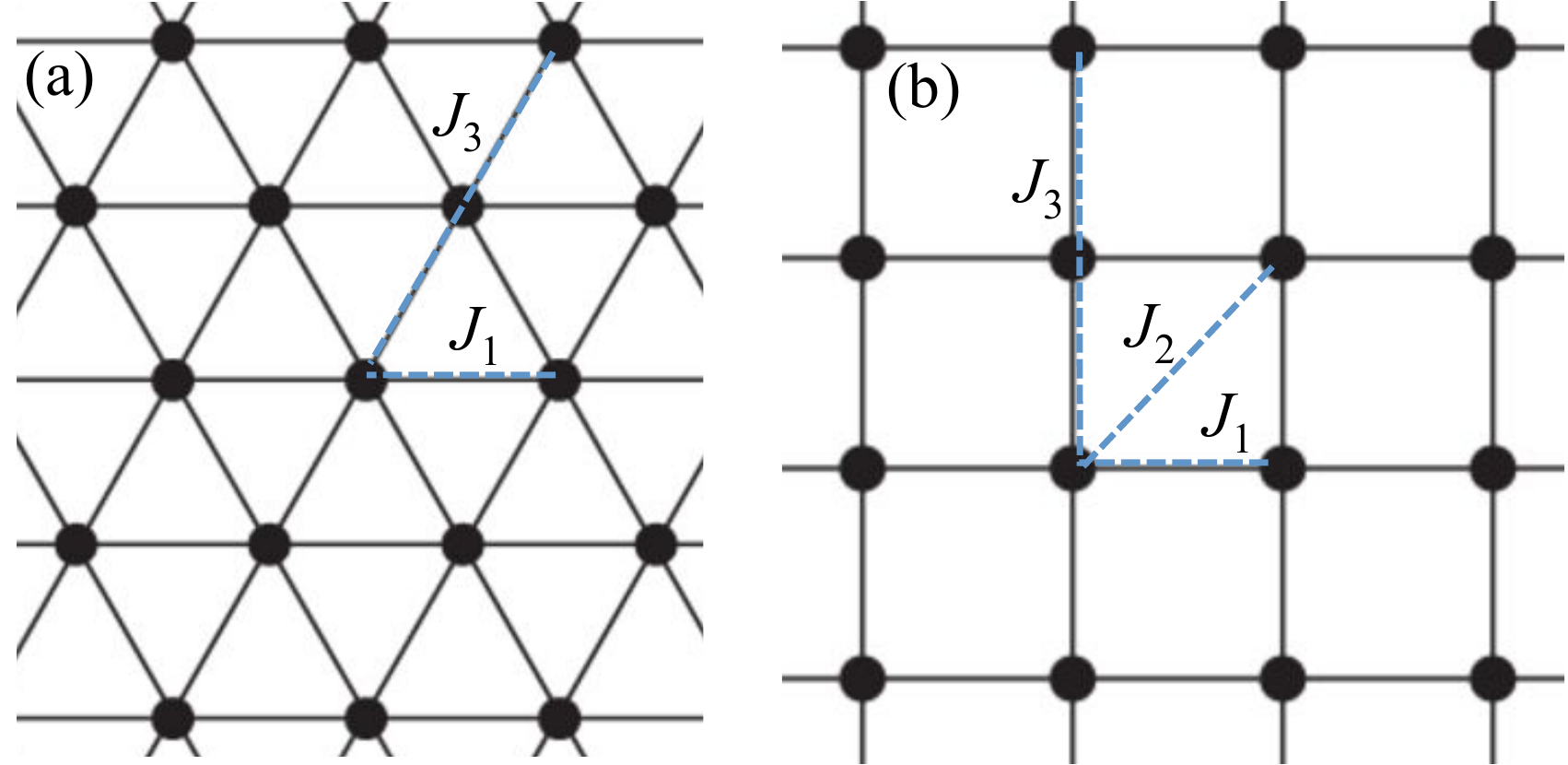,width=\columnwidth}
\caption{(color online) (a) $J_1$-$J_3$ classical Heisenberg model on a triangular lattice. (b) $J_1$-$J_2$-$J_3$ classical Heisenberg model on a square lattice.
} \label{f1}
\end{figure}

We present a derivation of Eq. \eqref{eqs2} using two frustrated Heisenberg models for the convenience of the following discussions. We consider the $J_1$-$J_3$ Heisenberg model on a triangular lattice [Fig.~\ref{f1} (a)] and the $J_1$-$J_2$-$J_3$ Heisenberg model on a square lattice [Fig.~\ref{f1} (b)]. The Hamiltonian generally can be written as
\begin{equation}\label{eqs3}
\mathcal{H}=-J_{ij}\sum _{\langle i,j \rangle}\mathbf{S}_i\cdot \mathbf{S}_j-\mathbf{H}_a \sum _i \mathbf{S}_i,
\end{equation}
where $J_{ij}=J_1$ is the nearest neighbor interaction, $J_{ij}=J_2$ is the next nearest neighbor and $J_{ij}=J_3$ is the next-next nearest neighbor interactions. In the Fourier space, Eq. \eqref{eqs3} becomes
\begin{equation}\label{eqs4}
\mathcal{H}=-\frac{N}{2}\int d q^3 J(\mathbf{q}) \mathbf{S}(\mathbf{q})\cdot\mathbf{S}(-\mathbf{q}),
\end{equation}
with $N$ the number of sites. For a triangle lattice the interaction $J(\mathbf{q})$ is
\begin{widetext}
\begin{equation}\label{eqs5}
J_{\triangle}(\mathbf{q})=2 J_1\left[\cos  \left(\frac{q_x}{2}+\frac{\sqrt{3} q_y}{2}\right)+\cos  \left(\frac{q_x}{2}-\frac{\sqrt{3} q_y}{2}\right)+\cos  q_x\right]+2 J_3\left[\cos  \left(q_x+\sqrt{3} q_y\right)+\cos  \left(q_x-\sqrt{3} q_y\right)+\cos  \left(2 q_x\right)\right],
\end{equation}
and for a square lattice
\begin{equation}\label{eqs6}
J_{\square}(\mathbf{q})=2 J_1\left[\cos(q_x) +\cos(q_y)\right]+2 J_2\left[\cos(q_x-q_y)  +\cos(q_x+q_y)  \right]+2 J_3\left[\cos(2q_x) + \cos (2q_y) \right],
\end{equation}
where $q$ is in unit of $1/a$. In the long wavelength limit when the optimal $Q_0$ that maximizes $J(\mathbf{q})$ is small, $Q_0 a \ll 1$, we expand $J(\mathbf{q})$ in $q$. We obtain for a triangular lattice
\begin{equation}\label{eqs7}
J_{\triangle}(\mathbf{q})\approx 6 \left(J_1+J_3\right)-\frac{3}{2} \left(J_1+4 J_3\right) q^2+\frac{3}{32} \left(J_1+16 J_3\right) q^4-\frac{1}{384} \left(J_1+64 J_3\right) q^6-\frac{\left(J_1+64 J_3\right) q^6 \cos (6 \phi )}{3840}+\mathcal{O}(q^8),
\end{equation}
and for a square lattice
\begin{equation}\label{eqs8}
J_{\square}(\mathbf{q})\approx 4 (J_1+J_2+J3)-(J_1+2 J_2+4 J_3) q^2+\frac{1}{16} (J_1+4 J_2+16 J_3)q^4+\frac{1}{48} (J_1-4 J_2+16 J_3)q^4 \cos (4 \phi ) +\mathcal{O}(q^6),
\end{equation}
\end{widetext}
with $q^2=q_x^2+q_y^2$ and $\tan\phi=q_y/q_x$. The spin lattice introduces a spatial anisotropy. For the triangular lattice, the spatial anisotropy is sixth order in $q$, $q^6 \cos 6\phi$. For a square lattice it is $q^4 \cos4\theta$. We can tune the parameters $J_1$, $J_2$, $J_3$ to ensure that the spatial anisotropy small to the order of $q^4$. Retaining terms up to quartic in $q$ and replacing $i q \rightarrow \nabla$, we obtain the effective Hamiltonian in Eq. \eqref{eqs2} after proper normalization of the length and energy.

\section{Single skyrmion solution}\label{Sec3}

In the previous studies, the skyrmion lattice in ISM with competing interactions is regarded as a superposition of triple helices \cite{PhysRevLett.108.017206,leonov_multiply_2015}. In the following sections, we emphasize the particle nature of skyrmions by exploring the skyrmion state in Eq. \eqref{eqs2} starting from a single skyrmion solution. First let us find a single skyrmion solution in the ferromagnetic background. Here we consider two dimensional systems with a perpendicular magnetic field. The solution is centrosymmetric and it is convenient to use the polar coordinate, $\mathbf{r}=(r, \phi)$. The spin vector can be represented as $\mathbf{S}=(\sin\theta\cos\varphi,\ \sin\theta\sin\varphi,\ \cos\theta)$. Here $\theta(r)$ only depends on $r$ and $\varphi=n \phi+\phi_0$ with integer $n$ the winding number (vorticity) and $\phi_0$ the helicity. For a skyrmion solution, $\theta(r=0)=\pi$ and $\theta(r=\infty)=0$. The skyrmion topological charge is
\begin{equation}\label{eqsb1}
N_s=\frac{1}{4\pi}\int dr^2\mathbf{S}\cdot(\partial_x\mathbf{S}\times\partial_y\mathbf{S})=-n.
\end{equation}
which is proportional to the vorticity and is independent of $\phi_0$. Because $\phi_0$ is associated with a smooth deformation of the spin texture, it does not affect the topology of the skyrmion. For convenience we also introduce the skyrmion topological charge density
\begin{equation}\label{eqsb1b}
\rho_s(\mathbf{r})=\frac{1}{4\pi}\mathbf{S}\cdot(\partial_x\mathbf{S}\times\partial_y\mathbf{S}).
\end{equation}

The total energy $E_T$ for a skyrmion solution with a winding number $n$ is
\begin{align}\label{eqsb2}
E_{T}=2\pi\int r dr\left[E_2+E_4-H_a \cos\theta\right],
\end{align}
\begin{align}\label{eqsb3}
E_2={ - \frac{1}{2}\left( {{{\left( {{\partial _r}\theta } \right)}^2} + \frac{{{n^2{\sin }^2}\theta }}{{{r^2}}}} \right)},
\end{align}
\begin{align}\label{eqsb4}
E_4 &=& \frac{1}{2}\left[ {{{\left( {{\partial _r}\theta } \right)}^4} + {{\left( {\partial _r^2\theta } \right)}^2}} \right] \nonumber\\
&+& \frac{{{\partial _r}\theta \left( { - r\sin \left( {2\theta } \right)  n^2+ 2{r^3}\partial _r^2\theta } \right) + n^4{{\sin }^2}\theta }}{{2{r^4}}} \nonumber\\
&+& \frac{1}{{2{r^2}}}\left[ {{{\left( {{\partial _r}\theta } \right)}^2}\left( {1 + 2 n^2 {{\sin }^2}\theta } \right) - n^2 \sin \left( {2\theta } \right)\partial _r^2\theta } \right].
\end{align}
The energy does not depend on $\phi_0$ and the sign of $n$, which means that skyrmions and antiskyrmions with different helicity have the same energy. Minimizing $E_T$ with respect to $\theta$, we obtain the equation for $\theta(r)$
\begin{align}\label{eqsb5}
\left(\frac{n^4}{2}-2n^2\right)\sin(2\theta)+r\left(1+2n^2\cos^2\theta \right) \partial_r\theta \nonumber \\
+r^2\left[\left((\partial_r\theta)^2-\frac{1}{2} \right)n^2 \sin(2\theta)-\left(1+2n^2\cos^2\theta \right)\partial_r^2\theta\right] \nonumber\\
+r^3\left(\partial_r\theta-2(\partial_r\theta)^3+2\partial_r^3\theta \right) \nonumber\\
+r^4\left(   H_a \sin\theta+\partial_r^2\theta-6(\partial_r\theta)^2\partial_r^2\theta+\partial_r^4\theta  \right)=0.
\end{align}
The equation is nonlinear near the center of skyrmion and generally requires numerical calculations. Let us consider the linear regime when $\theta\ll 1$ at $r\gg 1$. In this case, Eq. \eqref{eqsb5} can be simplified into
\begin{align}\label{eqsb6}
H_a \theta+\partial_r^2\theta+\partial_r^4\theta=0.
\end{align}
The solution can be written as
\begin{align}\label{eqsb7}
\theta(r)\sim \mathrm{Re}\left[c_1 K_0(q_+ r)+ c_2 K_0(q_- r)\right]  \nonumber \\
\sim \mathrm{Re}\left[c_1 \exp(-q_+ r)+ c_2 \exp(-q_- r)\right]/\sqrt{r},
\end{align}
at $r\gg 1$ with constants $c_1$ and $c_2$. Here $K_0(r)$ is the modified Bessel function of the second kind and $q_{\pm}$ are given by
\begin{align}\label{eqsb8}
q_{\pm}=\frac{{\sqrt { - 1 \pm \sqrt {1 - 4H_a } } }}{{\sqrt 2 }}.
\end{align} 
Note that the FM state is stable only at $H_a\ge H_s=1/4$, therefore $q_{\pm}$ is a complex number. Here $\mathrm{Re}[q_+]=\mathrm{Re}[q_-]$ describes the decay of $\theta$ and $\mathrm{Im}[q_+]=-\mathrm{Im}[q_-]$ describes the oscillation around zero. We may define $\xi\equiv 1/\mathrm{Re}[q_\pm]$ as the size of skyrmion. When $H_a$ approaches $H_s$ from above, $H_a-H_s=0^+$, the size of skyrmion $\xi$ diverges $\xi\sim 1/\sqrt{H_a-H_s}$ and $\mathrm{Im}[q_\pm]\approx 1/\sqrt{2}$, which is the optimal $Q_0$ for the conical spiral in the Hamiltonian Eq. \eqref{eqs2}. The size of skyrmion diverges at $H_s$, because there is a second order phase transition from the conical spiral to FM state at $H_s$ and therefore the length scale of the system diverges. This is different from the skyrmions stabilized by the DM interaction in two dimensions where the skyrmion size is always finite. On the other hand, when $\theta$ changes sign, the helicity of skyrmion also reverses sign. The helicity reversal was discussed in the $J_1$-$J_2$ Heisenberg model on a triangular lattice in Ref. \onlinecite{leonov_multiply_2015} and was observed experimentally in Ref. \onlinecite{Yu2012b}.

\begin{figure}[b]
\psfig{figure=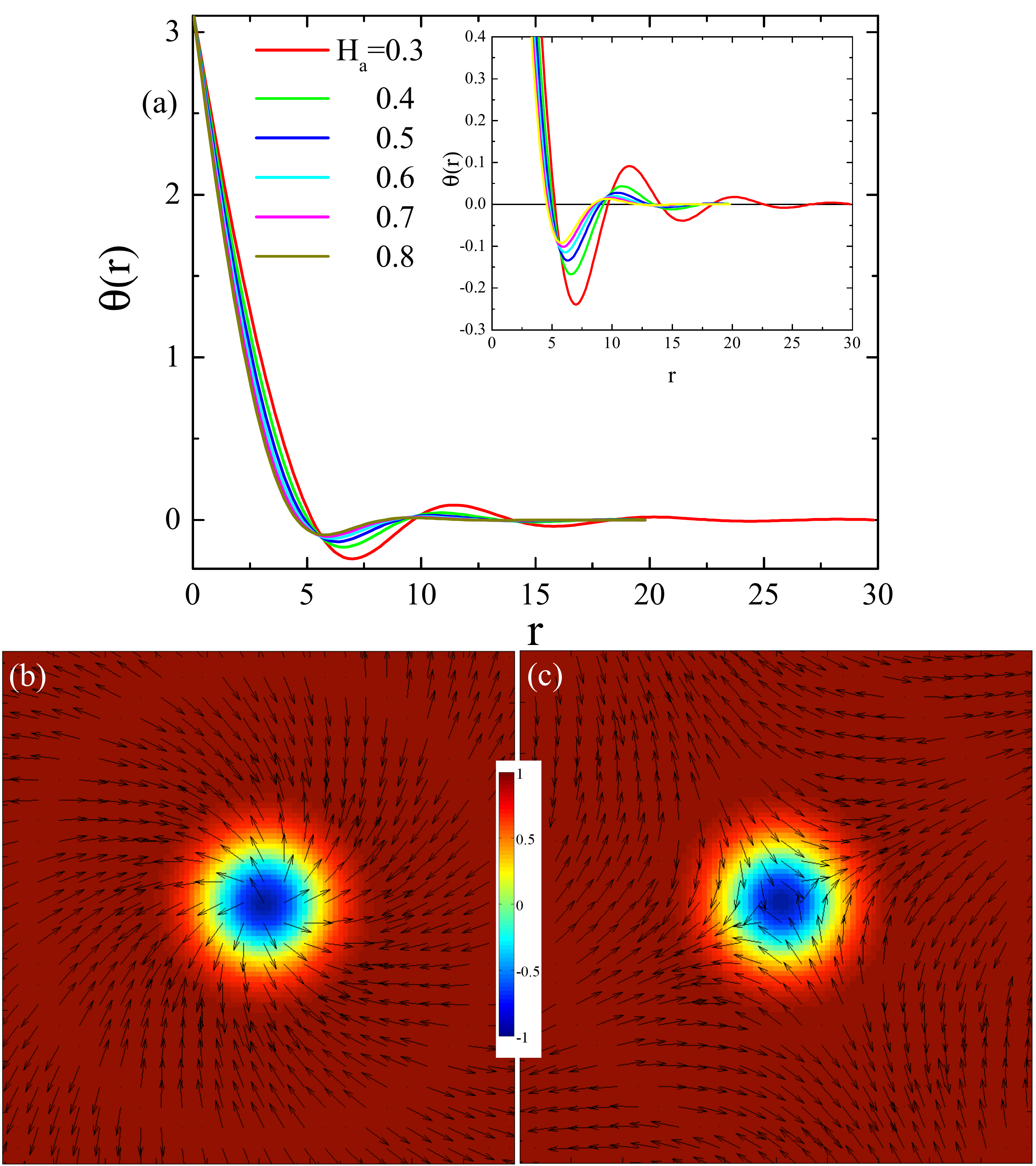,width=\columnwidth}
\caption{(color online) (a) Profile of $\theta(r)$ obtained by numerical  calculations of Eq. \eqref{eqsb5}. Inset shows that $\theta(r)$ decays and oscillates at large $r\gg 1$. Spin configurations for (b) an antiskyrmion with $N_s=-1$ and (c) a skyrmion with $N_s=1$. Color denotes the out-of-plane component of $\mathbf{S}$ and arrows represent the direction (not the magnitude) of the in-plane component. } \label{f2}
\end{figure}

In the presence of winding $n\neq 0$, the spin at the center $r=0$ must be parallel or antiparallel to field $S_z=\pm 1$. For $S_z=1$, the solution is topologically trivial $N_s=0$ and is connected smoothly to the FM state, therefore is an unstable solution. For $S_z=-1$, it is a skyrmion with $N_s=-n$. When one attempts to transform the skyrmion solution to the FM state, one has to flip the spin at the center of the skyrmion which results in a singular energy due to the winding of spins. Therefore the skyrmion solution is topologically protected and is a metastable solution with a finite lifetime. 

At a small $r$, we can expand $\theta(r)$ in $r$,  $\theta(r)=\pi+a_1 r+a_2 r^2+\mathcal{O}(r^3)$. For a skyrmion with a winding number $n=1$
\[
\theta(r)=\pi- a_1 r,
\]
 and $n=2$
 \[
\theta(r)=\pi- a_2 r^2.
\]
The results of $\theta(r)$ with $n=1$ obtained by numerical solution of Eq. \eqref{eqsb5} is shown in Fig. \ref{f2}. It oscillates and decays, which is consistent with the asymptotic behavior in Eq. \eqref{eqsb7}. Meanwhile the decay length increases when the field approaches the saturation field $H_s$. We may also define the core size $R_c$ of a skyrmion. The definition of $R_c$ is arbitrary and one may choose $\theta(r\le R_c)\le\pi/2$. The size of the nonlinear core depends weakly on magnetic field.

\begin{figure}[t]
\psfig{figure=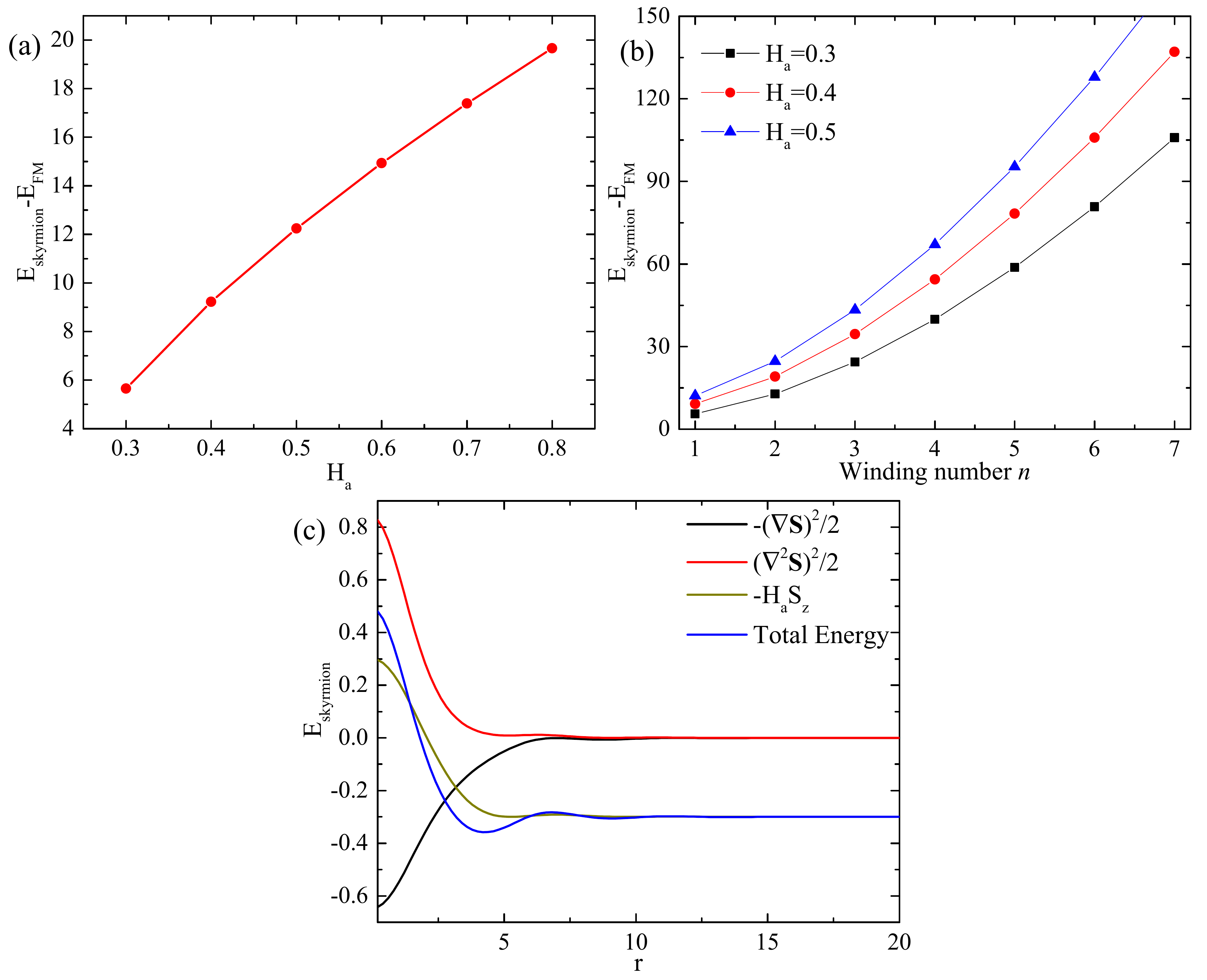,width=\columnwidth}
\caption{(color online) Energy of a skyrmion relative to the spin fully polarized state as a function (a) magnetic field $H_a$ and (b) winding number $n$. (c) Energy contribution of a skyrmion solution from different terms in Eq. \eqref{eqs2}.
} \label{f3}
\end{figure}

The energy of a skyrmion as a function of field $H_a$ and winding number $n$ is shown in Fig. \ref{f3}. The skyrmion has higher energy than the fully polarized state. The energy increases with $n$. The energy contributions from different terms in Hamiltonian Eq. \eqref{eqs2} as a function $r$ is displayed in Fig. \ref{f3}(c). The energy density is mainly contributed from the core region of a skyrmion. The term $-(\nabla \mathbf{S})^2$ gain energy and is balanced by the energy cost due to the term $(\nabla^2 \mathbf{S})^2$.

\section{Pairwise interaction between skyrmions}\label{Sec4}

In this section, we study the pairwise interaction between skyrmions. Because skyrmion and antiskyrmion are two degenerate solutions, we will consider the interaction both between two skyrmions, and between a skyrmion and an antiskyrmion. When two skyrmions are well separated, they interact through exchange of magnon excitations. Below we provide a linear theory to determine the mutual interaction between skyrmions. From this analysis, we can determine the the pairwise interaction as a function helicity and separation between skyrmions, but not the magnitude of the interaction. 

We introduce the magnon wave function
\begin{align}\label{eqs4_1}
\psi =S_x+i S_y,\ \ S_z=\sqrt{1-\psi \psi ^*}\approx 1-\frac{\psi \psi ^*}{2}.
\end{align}
The Hamiltonian in Eq. \eqref{eqs2} to the quadratic order in $\psi$ can be written as
\begin{align}\label{eqs4_2}
\mathcal{H}_{\psi }=\int dr^2 \left[H_a |\psi |^2-|\nabla \psi |^2+|\nabla ^2\psi |^2\right].
\end{align}
Here we have assumed that skyrmion lines are straight along the third direction and the problem reduces into two dimensions. The magnon is excited by the presence of skyrmions and we define a source $\tilde{F}(\mathbf{r}-\mathbf{R}_i)$ that depends on the nonlinear core of skyrmion, with $\mathbf{R}_i=(x_i,\ y_i)$ being the center of the skyrmion. Here $\tilde{F}(\mathbf{r}-\mathbf{R}_i)=0$ when $|\mathbf{r}-\mathbf{R}_i|\ge R_c$ with $R_c$ being an arbitrary radius separating the nonlinear core and linear tail of a skyrmion. Then the equation for $\psi$ is given by
\begin{align}\label{eqs4_3}
H_a \psi+\nabla^2\psi+\nabla^4\psi=\sum_i\tilde{F}(\mathbf{r}-\mathbf{R}_i).
\end{align}
Equation \eqref{eqs4_3} can be obtained by separating the nonlinear core and linear tail of skyrmion in Eq. \eqref{eqsb5}. The solution to Eq. \eqref{eqs4_3} in the region $|\mathbf{r}-\mathbf{R}_i|\gg R_c$ can be written as
\begin{align}\label{eqs4_3a}
\psi=\sum_i\psi_i\equiv\sum_i f(|\mathbf{r}-\mathbf{R}_i|)\exp[i (n_i\phi_i+\phi_{i0})],
\end{align}
\begin{align}\label{eqs4_3b}
f(r)=\mathrm{Re}[c_1K_n(q_+ r)+c_2K_n(q_-r)],
\end{align}
\begin{align}\label{eqs4_3c}
\phi_i=\arctan[(y-y_i)/(x-x_i)].
\end{align}
Here $\phi_{i0}$ is the helicity, $n_i$ is the winding number and the coefficients $c_1$ and $c_2$ are determined by $\tilde{F}(\mathbf{r}-\mathbf{R}_i)$. When skyrmions are well separated, $|\mathbf{r}-\mathbf{R}_i|\gg 2 R_c$, the contribution to the skyrmion interaction due to the interaction between the magnon and the nonlinear core is negligible, therefore it does not require to know the exact expression for $\tilde{F}(\mathbf{r}-\mathbf{R}_i)$ in the following calculations.

Excluding the nonlinear core contribution to the total energy, the magnon energy due to the linear tails of skyrmions becomes
\begin{align}\label{eqs4_4}
{E}_{\psi }=\int_{|\mathbf{r}-\mathbf{R}_i|>R_c} dr^2 [H_a |\psi |^2-|\nabla \psi |^2+|\nabla ^2\psi |^2] \nonumber\\
=\int_{|\mathbf{r}-\mathbf{R}_i|>R_c} dr^2{\psi ^*}\left( {H_a\psi  + {\nabla ^2}\psi  + {\nabla ^4}\psi } \right)\nonumber\\
+\int_{|\mathbf{r}-\mathbf{R}_i|>R_c} dr^2[ \nabla \left( {\nabla {\psi ^*}{\nabla ^2}\psi } \right) - \nabla \left( {{\psi ^*}{\nabla ^3}\psi } \right)- \nabla \left( {{\psi ^*}\nabla \psi } \right)  ]\nonumber\\
=\int_{|\mathbf{r}-\mathbf{R}_i|>R_c} dr^2{\psi ^*}\sum_i\tilde{F}(\mathbf{r}-\mathbf{R}_i) \nonumber\\
+\oint_{|\mathbf{r}-\mathbf{R}_i|>R_c} d\mathbf{l}\cdot[{\nabla {\psi ^*}{\nabla ^2}\psi } - {{\psi ^*}{\nabla ^3}\psi } -  {{\psi ^*}\nabla \psi }  ],
\end{align}
where $\oint_{|\mathbf{r}-\mathbf{R}_i|>R_c}  d\mathbf{l}$ is integration around a circle with a radius $R_c$ around the skyrmion centers and the direction of $\mathbf{l}$ is normal to the circle. The other contribution at $\oint_{|\mathbf{r}-\mathbf{R}_i|=\infty} d\mathbf{l}$ vanishes because the magnon wave function decays exponentially. Neglecting the nonlinear core contribution, the magnon energy can be written as
\begin{align}\label{eqs4_5}
{E}_{\psi }=\oint_{|\mathbf{r}-\mathbf{R}_i|>R_c} d\mathbf{l}\cdot[{\nabla {\psi ^*}{\nabla ^2}\psi } - {{\psi ^*}{\nabla ^3}\psi } -  {{\psi ^*}\nabla \psi }  ].
\end{align}
We then calculate ${E}_{\psi }$ in the presence of two skyrmions. We have for ${E}_{\psi }$ 
\begin{align}\label{eqs4_6}
{E}_{\psi }=2E_s+E_{12},
\end{align}
with $E_s$ the self-energy of a skyrmion
\begin{align}\label{eqs4_6a}
{E}_{s }=\oint_{|\mathbf{r}-\mathbf{R}_i|>R_c} d\mathbf{l}\cdot[{\nabla {\psi_i ^*}{\nabla ^2}\psi_i } - {{\psi_i ^*}{\nabla ^3}\psi_i } -  {{\psi_i ^*}\nabla \psi_i }  ],
\end{align}
and $E_{12}$ the pairwise interaction
\begin{align}\label{eqs4_6b}
E_{12}=\oint_{|\mathbf{r}-\mathbf{R}_i|>R_c} d\mathbf{l}\cdot[{\nabla {\psi_1 ^*}{\nabla ^2}\psi_2 } - {{\psi_1 ^*}{\nabla ^3}\psi_2 } -  {{\psi_1 ^*}\nabla \psi_2 }  +(1\leftrightarrow  2)],
\end{align}
where $\psi_i$ is the magnon induced by the skyrmion at $\mathbf{R}_i$ defined in Eq. \eqref{eqs4_3a}.

\begin{figure}[t]
\psfig{figure=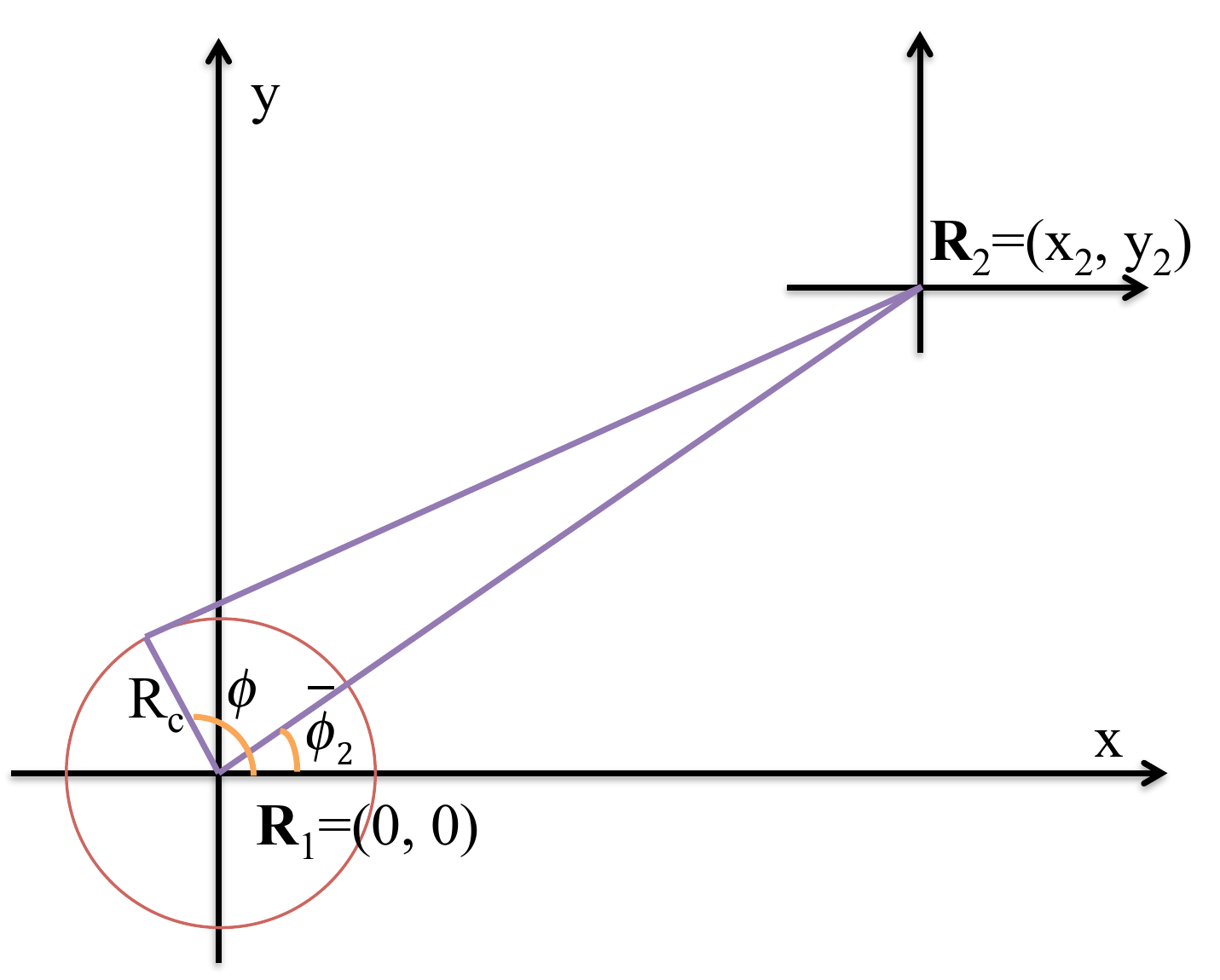,width=\columnwidth}
\caption{(color online) Coordinate system used in the evaluation of the pairwise interaction between skyrmions and antiskyrmions. $\mathbf{R}_i$ is the skyrmion center and the circle with radius $R_c$ is the integration contour.
} \label{f4a}
\end{figure}

To evaluate ${E}_{\psi }$ we take $E_{\psi,1}=\oint_{|\mathbf{r}-\mathbf{R}_i|>R_c}  d\mathbf{l}\cdot[{\psi_1 ^*}\nabla \psi_2 +{\psi_2 ^*}\nabla \psi_1]$ as an example and other terms can be evaluated similarly,
\begin{align}\label{eqs4_7}
{E}_{\psi,1 }=\oint_{|\mathbf{r}-\mathbf{R}_1|>R_c}  d\mathbf{l}\cdot[  {{\psi_1 ^*(\mathbf{R}_c)}\nabla \psi_2(\mathbf{r}-\mathbf{R}_2) } ]\nonumber\\
+\oint_{|\mathbf{r}-\mathbf{R}_2|>R_c}  d\mathbf{l}\cdot[  {{\psi_1 ^*(\mathbf{r}-\mathbf{R}_1)}\nabla \psi_2(\mathbf{R}_c) } ]+c.c.
\end{align}

We choose the coordinate system shown in Fig. \ref{f4a}. In the limit $R_2\gg R_c$, we take the approximation for $\psi_2(\mathbf{r}-\mathbf{R}_2)$
\begin{align}\label{eqs4_8}
\psi_2(\mathbf{r}-\mathbf{R}_2)\approx \nonumber\\
[{f(R_2) - \partial_r f(R_2)r\cos( \phi  - \bar{\phi} _2 )}]\exp[i\left( n_2 ({{\bar{\phi} _2} + \pi })  + {\phi _{20}} \right)],
\end{align}
\begin{align}\label{eqs4_9}
\bar{\phi}_2=\arctan(y_2/x_2),
\end{align}
where we have also used $\partial_r f(R_2)\ll f(R_2)/R_2$. We then obtain for two skyrmions $n_1=n_2=1$
\begin{align}\label{eqs4_10}
{E}_{\psi,1 }=-2\pi R_c\partial_rf(R_{12})[f(R_c)+R_c\partial_r f(R_c)]\cos(\phi_{20}-\phi_{10}),
\end{align}
and for skyrmion and antiskyrmion, $n_1=1$, $n_2=-1$
\begin{align}\label{eqs4_11}
{E}_{\psi,1 }=-2\pi R_c\partial_rf(R_{12})[f(R_c)+R_c\partial_r f(R_c)]\cos(\phi_{20}-\phi_{10}-2\phi_{12}),
\end{align}
with $R_{12}\equiv |\mathbf{R}_1-\mathbf{R}_2|$ and $\phi_{12}\equiv \arctan[(y_2-y_1)/(x_2-x_1)]$. Here we are interested in the dependence of pairwise interaction on $R_{12}$ and $\phi_{i0}$ and we do not write the full expression. For two skyrmions, the dependence of the interaction on separation and helicity is given by
\begin{align}\label{eqs4_12}
{E}_{12 }(R_{12})\sim \mathrm{Re}[\exp(-q_\pm R_{12})] \cos(\phi_{20}-\phi_{10}),
\end{align}
and for a skyrmion and an antiskyrmion, it is given by
\begin{align}\label{eqs4_13}
{E}_{12 }(R_{12})\sim \mathrm{Re}[\exp(-q_\pm R_{12})]\cos(\phi_{20}-\phi_{10}-2\phi_{12}).
\end{align}
We introduce $\bar{\phi}_i(\mathbf{R}_j-\mathbf{R}_i)$ as the phase of the skyrmion at $\mathbf{R}_j$ for a skyrmion located at $\mathbf{R}_i$. The interaction between skyrmions/antiskyrmions can be written as
\begin{align}\label{eqs4_14}
{E}_{12 }(R_{12})\sim \exp(-q_\pm R_{12})\cos[\bar{\phi}_{1}(\mathbf{R}_2-\mathbf{R}_1)-\bar{\phi}_{2}(\mathbf{R}_1-\mathbf{R}_2)].
\end{align}
The interaction between skyrmions depends on their relative helicity and their separation $R_{12}$. Meanwhile the interaction between a skyrmion and an antiskyrmion depends on angle $\phi_{12}$ in addition to the relative helicity. The interaction is not isotropic and it depends the relative orientation between the skyrmion and antiskyrmion with respect to their helicity.

The dependence of $E_{12}$ on helicity and $R_{12}$ can be understood as follows. The interaction range is determined by the magnon gap. For $H_a>H_s$, the interaction is short ranged with a decay length $1/\mathrm{Re}[q_\pm]$ in Eq. \eqref{eqsb8}. The interaction range increases when $H_a$ approaches $H_s$ and finally diverges at $H_s$. Since the magnon wave function oscillates as a function of distance, the interaction is nonmonotonic as a function of distance between skyrmions. For a certain separation, there is cancellation of the magnon excitation which corresponds to a local minimum in the pair potential. If there is addition of the magnon excitation, it corresponds to a local maximum in the pair potential. Because the phase of the magnon excitation depends on helicity of skyrmion, if we reverse the helicity of a skyrmion, then the original local minimum in the pair potential because a local maximum and vice versa. When the separation between two skyrmions is comparable to their size, the interaction is induced by overlapping their nonlinear cores. In this nonlinear regime, there are many-body interactions among skyrmions if we consider the case with many skyrmions.
 
\begin{figure}[t]
\psfig{figure=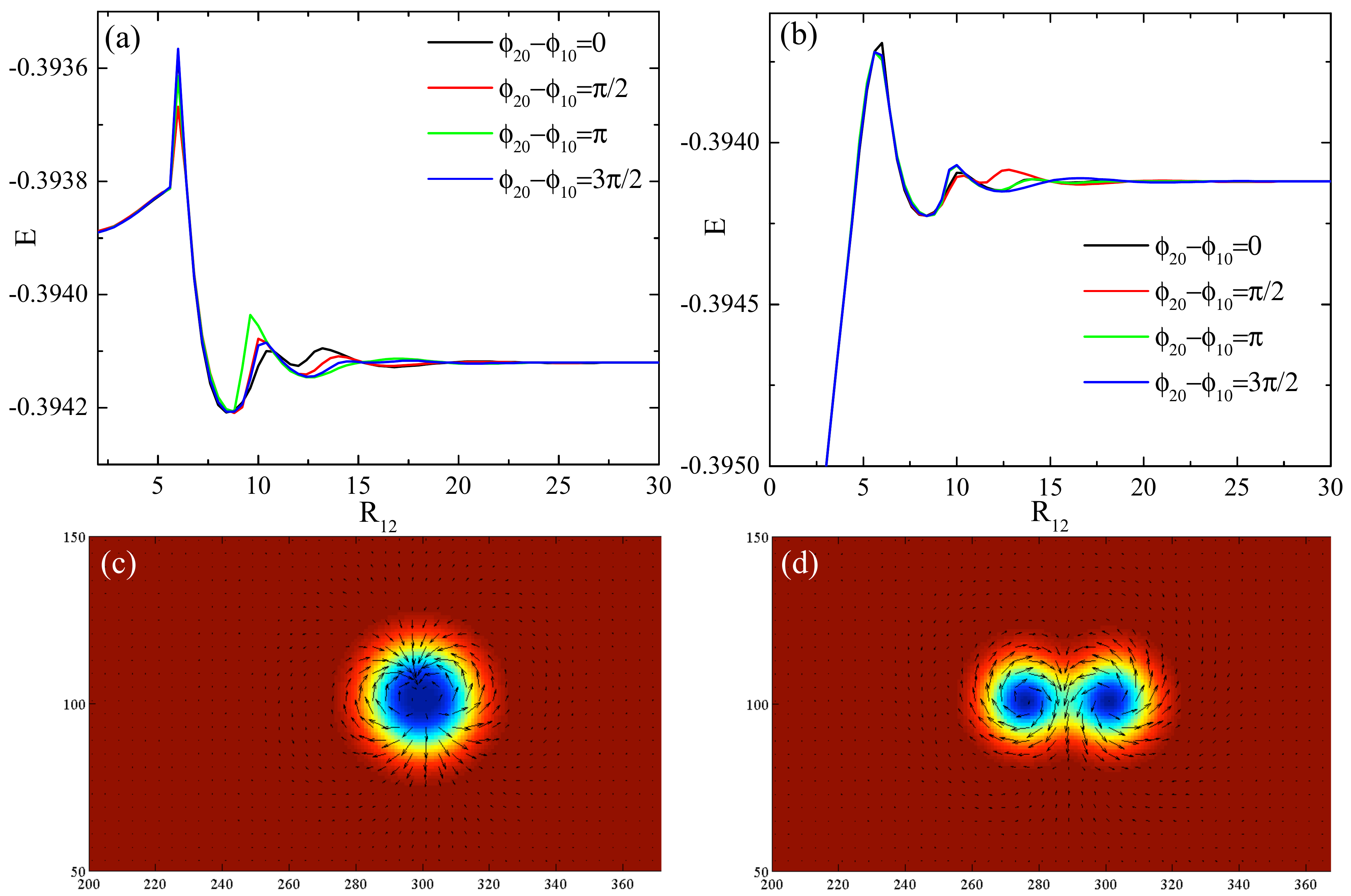,width=\columnwidth}
\caption{(color online) Energy of (a) two skyrmions and (b) a skyrmion and an antiskyrmion with a separation $R_{12}$ and relative helicity $\phi_{20}-\phi_{10}$. In (b) we set $\phi_{12}=0$. (c) is the spin configuration for a skyrmion with topological charge $N_s=-2$ and (d) is the spin configuration for a skyrmion and an antiskyrmion when they approach to each other.
} \label{f4}
\end{figure}

We calculate numerically the pairwise interaction between two skyrmions as a function of separation $R_{12}$ and compare to the linear analysis. We fix the spin at the center of the skyrmions, which effectively pin the skyrmion at a desired position. We then relax the system according to the Landau-Liftshitz-Gilbert dynamics and obtain the stationary energy $E$. The results are shown in Fig. \ref{f4}. For a large separation, the interaction oscillates and decays. The interaction also depends on the relative helicity $\phi_{20}-\phi_{10}$ between skyrmion/antiskyrmion. For a $\phi_{20}-\phi_{10}=\pi$ phase shift in the helicity, it turns attraction for $\phi_{20}-\phi_{10}=0$ into repulsion and vice versa, which is consistent with the linear analysis in Eqs. \eqref{eqs4_12} and \eqref{eqs4_13}. At a small separation, we cannot identify two independent skyrmions because their cores start to overlap. Therefore the relative helicity is no longer well defined. According to the calculations, $E(R_{12})$ for different helicity collapse into a single curve. When $R_{12}=0$, the two skyrmions merge into a skyrmion with $N_s=-2$ or giant skyrmion [Fig. \ref{f4} (c)], while in the case of skyrmion and antiskyrmion, they annihilate and no skyrmion is left in the system  [Fig. \ref{f4} (d)]. Note that the magnetization is not uniform in space because we fix the spin antiparallel to the field at the center. To merge into a giant skyrmion, the system needs to overcome a steep energy barrier as presented in Fig. \ref{f4} (a) and (b). Similar situation occurs for annihilating a skyrmion and an antiskyrmion.  

\section{Creation of skyrmions by annealing}\label{Sec5}

\begin{figure}[t]
\psfig{figure=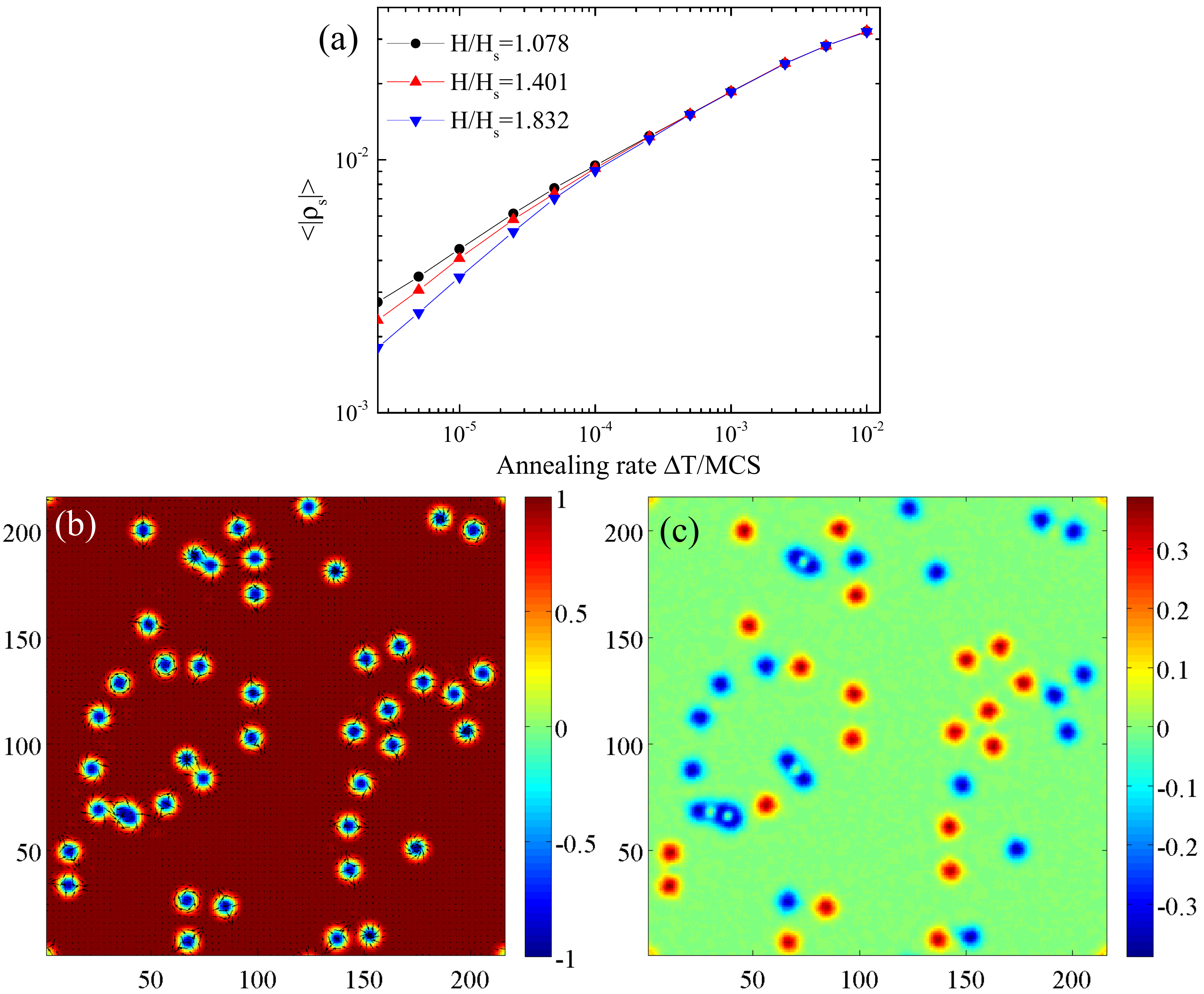,width=\columnwidth}
\caption{(color online) (a) The averaged absolute skyrmion charge density, $\langle |\rho_s| \rangle=\frac{1}{4\pi L^2}\int dr^2|\mathbf{S}\cdot(\partial_x\mathbf{S}\times\partial_y\mathbf{S})|$ with $L$ the system size as a function of annealing rate, where $\langle\cdots\rangle$ denotes the average of independent annealing process. To obtain a better statistics, the results are obtained by averaging over 20 independent runs with different initial configurations. (b) and (c) correspond to the spin profile and skyrmion topological charge density $\rho_s$ respectively obtained after annealing with annealing rate $\Delta T=0.00001 J_1$ per Monte Carlo Sweep (MCS) at $H_a/H_s=1.617$. The results are obtained by Monte Carlo annealing in the $J_1$-$J_2$-$J_3$ model on a square lattice. Here we take $Q_0=2\pi/18$ and $J_1-4J_2+16J_3=0$. The saturation field is $H_s=0.001856J_1$.
} \label{f5}
\end{figure}

Since a skyrmion is a metastable state in the ferromagnetic state, one natural way to excite skyrmions is by annealing. In Fig. \ref{f5} we present the results obtained by annealing in Monte Carlo simulations of lattice model Eq. \eqref{eqs3} in the long wavelength limit. Initially the system is equilibrated at high $T$ and is in the paramagnetic state. Then we gradually reduce temperature with a rate $\Delta T$ per every Monte Carlo sweep (MCS). In this process, skyrmions and antiskyrmions are nucleated according to the Kibble-Zurek mechanism \cite{Kibble1976,Zurek1985}. Because the Hamiltonian Eq. \eqref{eqs2} does not distinguish between skyrmions and antiskyrmions, they are created equal therefore the total topological charge is zero after averaging over the same annealing process. Since the interaction between skyrmions and antiskyrmions is nonmonotonic as a function of separation, and since there is a steep energy barrier for the annihilation between skyrmions and antiskyrmions, they are trapped by the local minimum in their interaction potential at low temperatures and they do not annihilate. The density of the skyrmions can be controlled by the annealing rate [Fig. \ref{f5}(a)]. At the initial state when $T\gg H_a$, the absolute of skyrmion density does not depend on magnetic field. For a fast annealing, the final state resembles the initial state therefore the skyrmion density almost does not depend on magnetic field. For a slow annealing, the system can reach a lower energy state by reducing the skyrmion density because excitation of skyrmions costs energy. In this region, because the skyrmion energy increases with magnetic field [Fig. \ref{f3}], the skyrmion density decreases with field.  In the presence of random pinning potential produced by defects, skyrmions and antiskyrmions can be trapped by the pinning potential, which facilitates the creation of a metastable skyrmion state by annealing.  

\section{Stabilizing Skyrmion lattice by an easy-axis anisotropy}\label{Sec6}

A single skyrmion in the ferromagnetic background is a metastable state. A general question is whether we can stabilize the skyrmion lattice as a thermodynamically stable phase. The triangular lattice of skyrmion can be regarded as a superposition of three helices with ordering wave vector rotated by $120^{\circ}$. Therefore the spatial anisotropy introduced by underlying triangular lattice of spin favors the triple-$Q$ ordering. The spatial anisotropy is proportional to $q^6$ therefore a large wavevector $Q$ (or short period of helix) is preferred to stabilize the triangular lattice of skyrmion. The skyrmion lattice discussed in $J_1$-$J_3$ or $J_1$-$J_2$ model in Ref. \onlinecite{PhysRevLett.108.017206} is only stabilized for $Q>Q_c\sim 1$. \cite{Satoru2015b}

\begin{figure}[t]
\psfig{figure=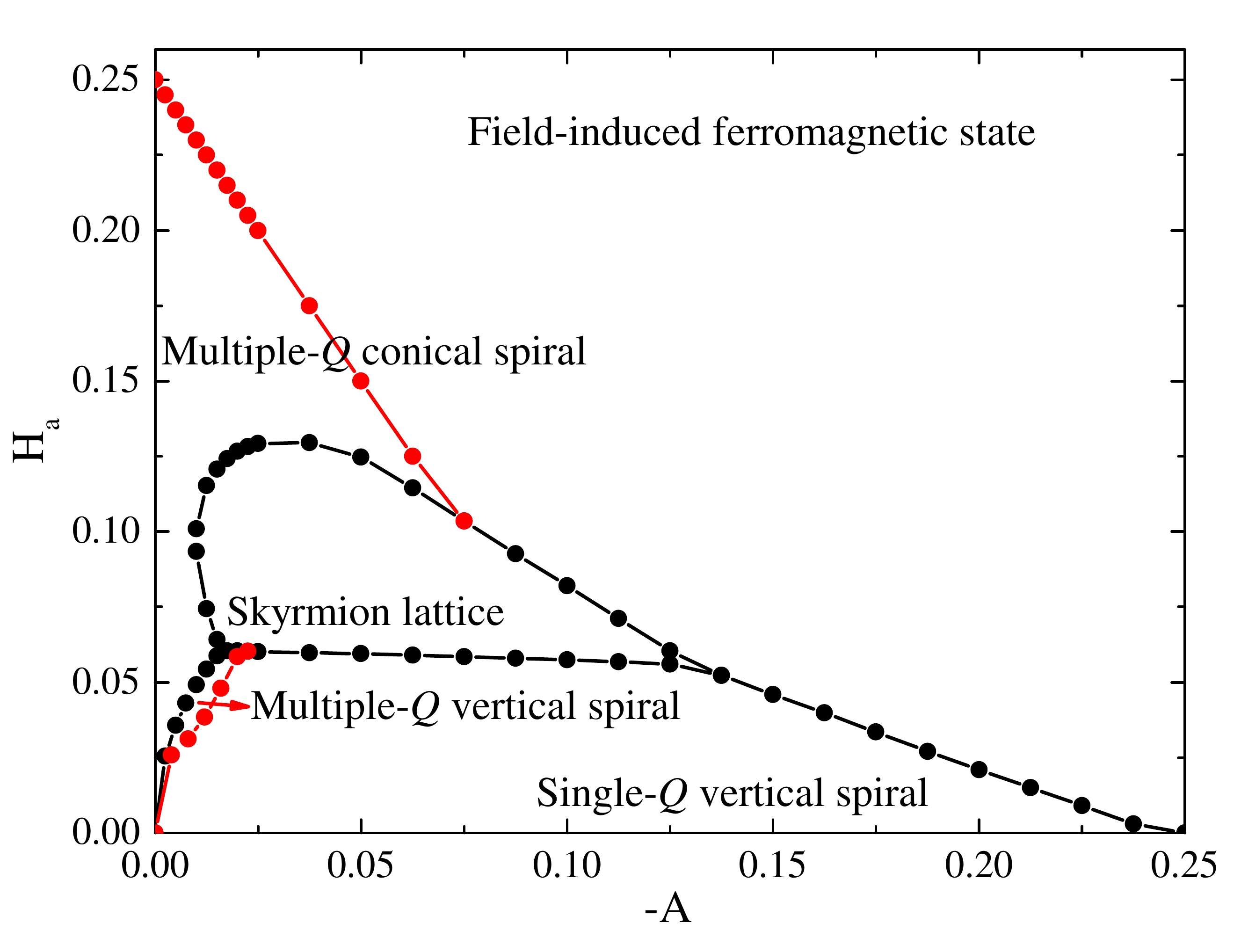,width=\columnwidth}
\caption{(color online) Phase diagram obtained by variational calculations of Eq. \eqref{eqsd1} at $T=0$. Black lines represent the first order phase transition and the red lines denote the second order phase transition.
} \label{f6}
\end{figure}

As far as the skyrmion size is much bigger than the lattice parameter of the spin system, or in the long wavelength limit as considered here, the spin lattice becomes irrelevant. Here we consider the stabilization of skyrmion lattice by a perpendicular easy axis anisotropy. In the case of skyrmions stabilized by the DM interaction, it has already been demonstrated that an easy axis anisotropy favors the skyrmion lattice. \cite{Butenko2010,Wilson2012,Huang2012} For frustrated systems, this was demonstrated in the $J_1$-$J_2$ Heisenberg model on a triangular lattice. \cite{leonov_multiply_2015} The reason is that in the skyrmion lattice, the majority of spin is along the easy axis thus has lower energy in comparison to that of helix phase. We remark that the easy axis anisotropy does not modify qualitatively the properties of a single skyrmion discussed in Sec. \ref{Sec3}. The spin Hamiltonian in the presence of an easy axis anisotropy in two dimensions is
\begin{equation}\label{eqsd1}
\mathcal{H}=\int dr^2\left[-\frac{1}{2} (\nabla \mathbf{S})^2+\frac{1}{2}(\nabla^2 \mathbf{S})^2-\mathbf{H}_a\cdot\mathbf{S}+A S_z^2\right],
\end{equation}
with $A<0$. First we study the phase diagram of Eq. \eqref{eqsd1} at $T=0$ by variational calculations. We consider the following six states:

1. A spin fully polarized state, $\mathbf{S}=(0, 0, 1)$. The energy density is $E_{\mathrm{FM}}=-H_a+A$.

2. Single-$Q$ conical spiral: in this state, the transverse components of spin rotate in a plane parallel to the ordering wave vector $\mathbf{Q}_0$ and the amplitude of the longitudinal component is constant. The ansatz for the spin state is  $\mathbf{S}=[\sin\theta\cos(\mathbf{Q}_0\cdot \mathbf{r}),\ \sin\theta\sin(\mathbf{Q}_0\cdot \mathbf{r}),\ \cos\theta]$ when $H_a\le H_s$ with $H_s=(2A+1/4)$ being the saturation field. The canting angle $\theta$ is given by $\cos\theta=H_a/H_s$. The corresponding energy is $E_c=-(H_s+4H_a^2)/8 H_s$. The conical spiral is no long stable when $A\le -1/8$.

3. Single-$Q$ vertical spiral: in this state, the spins rotate in a plane parallel to the magnetic field. The direction of $\mathbf{Q}_0$ is not determined. Here we take $\mathbf{Q}_0$ perpendicular to the spin rotation plane. The ansatz for the spin state is $\mathbf{S}=\mathbf{m}/|m|$ with $\mathbf{m}=[0,\ a_1\sin(q_v x),\ a_2\cos(q_v x)+\bar{m}]$, where we have assumed $q_v$ is along the $x$ direction. Here $a_1$, $a_2$, $q_v$ and $\bar{m}$ are variational parameters.

4. Triangular lattice of skyrmion: the spin ansatz can be written as $\mathbf{S}=\mathbf{m}/|m|$ with 
\[
m_x=a_1\left(-\frac{\sqrt{3}}{2} \sin(\mathbf{q}_2\cdot \mathbf{r})+\frac{\sqrt{3}}{2} \sin(\mathbf{q}_3\cdot \mathbf{r})\right),
\]
\[
m_y=a_1\left(\sin(\mathbf{q}_1\cdot\mathbf{r})-\frac{1}{2} \sin(\mathbf{q}_2\cdot \mathbf{r})-\frac{1}{2} \sin(\mathbf{q}_3\cdot \mathbf{r})\right),
\]
\[
m_z=-a_2\left(\cos(\mathbf{q}_1\cdot\mathbf{r})+ \cos(\mathbf{q}_2\cdot \mathbf{r})+ \cos(\mathbf{q}_3\cdot \mathbf{r})\right)+\bar{m}.
\]
Here $\mathbf{q}_1=q_v\hat{x}$, $\mathbf{q}_2=(-\frac{1}{2}\hat{x}+\frac{\sqrt{3}}{2}\hat{y})q_v$ and $\mathbf{q}_3=(-\frac{1}{2}\hat{x}-\frac{\sqrt{3}}{2}\hat{y})q_v$ with $\hat{x}$ and $\hat{y}$ unit vectors along the $x$ and $y$ directions respectively. The Hamiltonian Eq. \eqref{eqs2} has $U(1)$ symmetry. In the skyrmion lattice phase, the $U(1)$ symmetry is broken spontaneously by taking one arbitrary helicity for all skyrmions. In the ansatz, we have chosen the helicity to be $\phi_0=\pi/2$. The variational parameters are $a_1$, $a_2$, $q_v$ and $\bar{m}$.

5. Multiple-$Q$ conical spiral: it was discussed in Ref. \onlinecite{leonov_multiply_2015} that the single-$Q$ conical spiral is unstable with respect to an easy axis anisotropy by developing modulation in other directions. This is an example that an easy axis anisotropy prefers to stabilize multiple-$Q$ states. The ansatz for the spin state is $\mathbf{S}=\mathbf{m}/|m|$ with 
\[
m_x=a_1 \cos(\mathbf{q}_1\cdot \mathbf{r})+ a_2\cos(\mathbf{q}_2\cdot \mathbf{r}),
\]
\[
m_y=-a_1\sin(\mathbf{q}_1\cdot \mathbf{r})+ a_2 \sin(\mathbf{q}_2\cdot \mathbf{r}),
\]
\[
m_z=a_3\cos(\mathbf{q}_3\cdot \mathbf{r})+\bar{m}.
\]
By introducing additional modulation in $m_z$, the system gains energy in the easy axis anisotropy, which outweighs the energy costs due to higher harmonics. This multiple-$Q$ conical spiral can be further classified according to whether $a_1=a_2$ or not. \cite{leonov_multiply_2015} Here we generally refer them as multiple-$Q$ conical spirals.

6. Multiple-$Q$ vertical spiral: the single-$Q$ vertical spiral is also found to be unstable by developing additional modulation in certain range of $A$ in Ref. \onlinecite{leonov_multiply_2015}. This multiple-$Q$ vertical spiral can be described by
\[
m_x=a_1 \cos(\mathbf{q}_2\cdot \mathbf{r}+\phi_v)-a_1 \cos(\mathbf{q}_3\cdot \mathbf{r}-\phi_v),
\]
\[
m_y=-a_2\sin(\mathbf{q}_1\cdot \mathbf{r}),
\]
\[
m_z=a_2\cos(\mathbf{q}_1\cdot \mathbf{r})+\bar{m}.
\]
Here we have chosen the spin rotation plane to be perpendicular to the dominant ordering wave vector.
 
\begin{figure}[t]
\psfig{figure=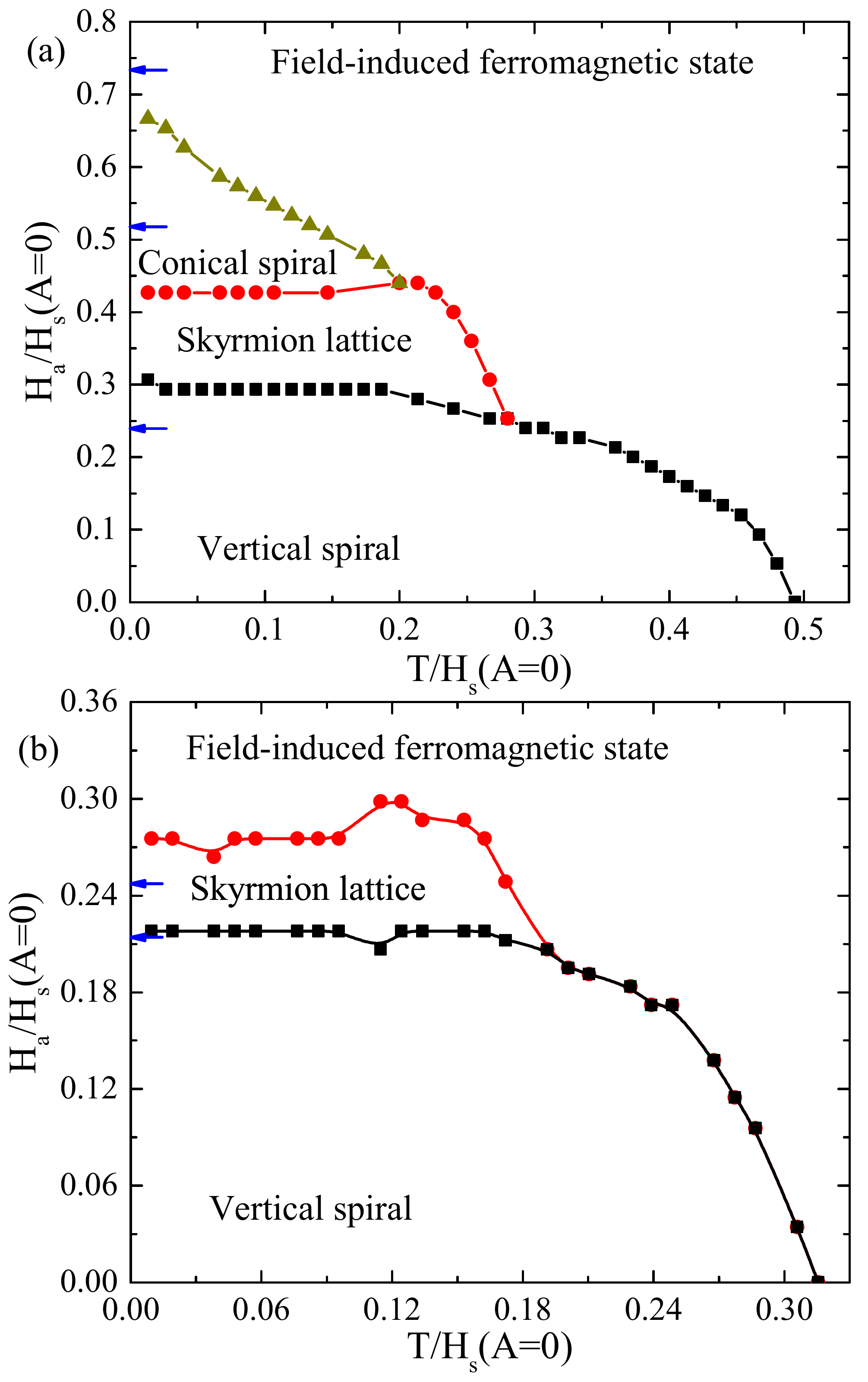,width=\columnwidth}
\caption{(color online) Phase diagram obtained by Monte Carlo simulations of (a) the $J_1$-$J_2$-$J_3$ Heisenberg model on a square lattice and (b) the $J_1$-$J_3$ Heisenberg model on a triangular lattice. Here for (a) $J_3=-0.15J_1$, $J_1-4J_2+16J_3=0$ and $A=-2 H_s(A=0)/15$ with the saturation field at $A=0$, $H_s(A=0)=0.15 J_1$; for (b) $J_3=-0.5 J_1$ and $A=-0.478 H_s(A=0)$ with $H_s(A=0)=1.045 J_1$. Arrows denote the phase boundaries at $T=0$ obtained by variational calculation in Fig. \ref{f6}.
} \label{f7}
\end{figure}

\begin{figure*}[t]
\psfig{figure=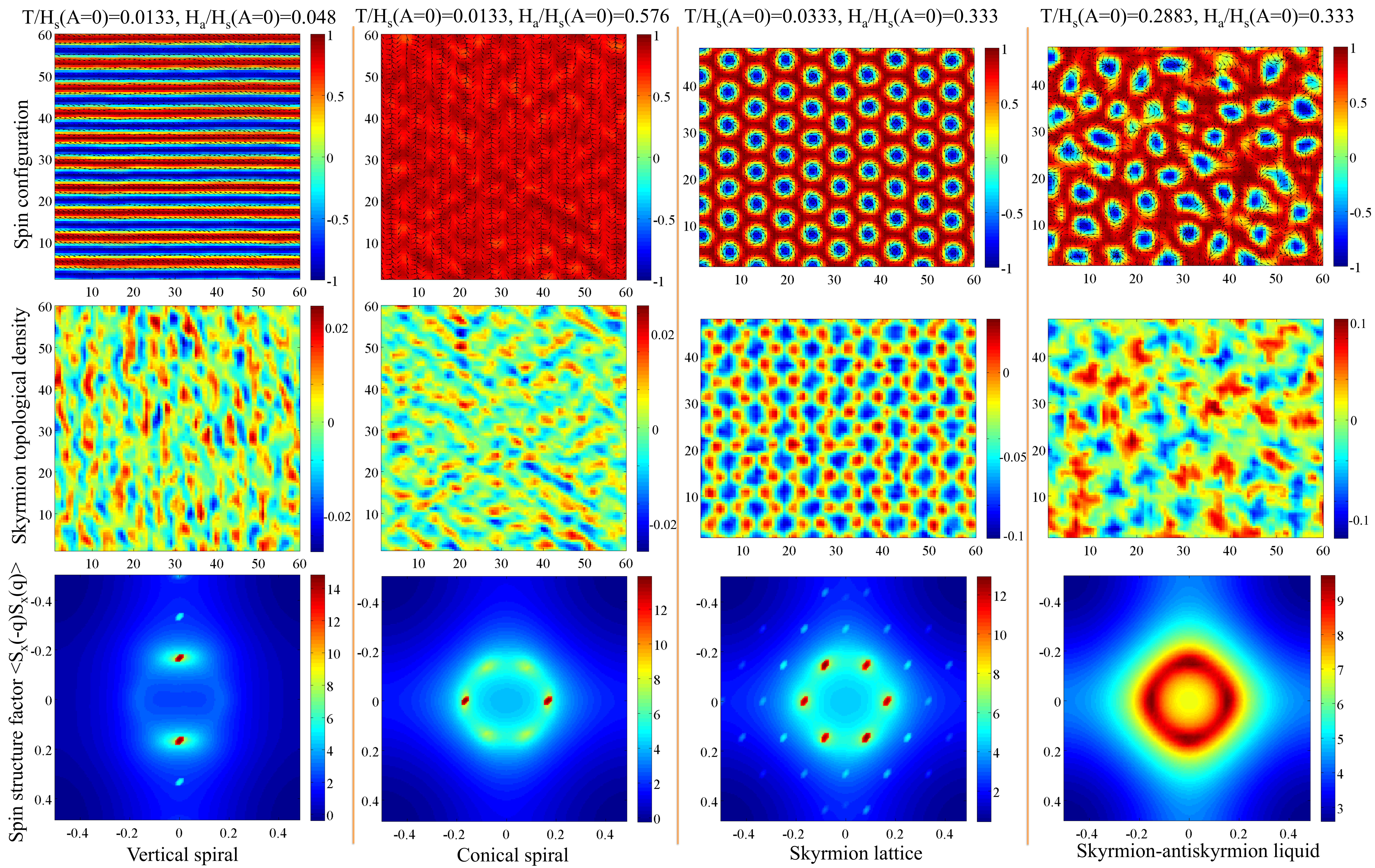,width=18cm}
\caption{(color online) Spin configuration (first row), skyrmion topological charge density $\rho_s$ (second row) and spin structure factor $\log(\langle S_x(\mathbf{q})S_x(-\mathbf{q})\rangle)$ (third row) for the vertical spiral (first column), conical spiral (second column), skyrmion lattice (third column) and skyrmion-antiskyrmion liquid (fourth column). The results are obtained by Monte Carlo simulations of the $J_1$-$J_2$-$J_3$ Heisenberg model on a square lattice and parameters are the same as those in Fig. \ref{f7}. In the skyrmion-antiskyrmion liquid phase, the structure factor is not a perfect ring because of the spatial anisotropy introduced by the spin lattice.
} \label{f8}
\end{figure*}

Except for the single-$Q$ conical spiral, all the other states have harmonics because of the constraint $|S|=1$. We use relaxation method to find the lowest energy state in the variational space, $\partial_t a_v=-\partial E(a_v)/\partial a_v$, for all the variational parameters $a_v$. We then compare the energy among the these states. The phase diagram at $T=0$ is shown in Fig. \ref{f6}. All the phase transitions are of the first order except for the transition between the multiple-$Q$ conical spiral and ferromagnetic state, and the transition between the single-$Q$ vertical spiral and multiple-$Q$ vertical spiral. There are three notable features. First, a skyrmion lattice is stabilized in the intermediate magnetic field in the presence of an easy axis anisotropy. It could be skyrmion or antiskyrmion lattice, thus breaks the $Z_2$ symmetry. Secondly the single-$Q$ conical spiral at $A=0$ is unstable with respect to an infinitesimal $A$ and the multiple-$Q$ conical spiral is stabilized. Thirdly the easy axis anisotropy stabilizes a vertical spiral at low field with respect to the conical spiral. From Fig. \ref{f6}, it is clear that an easy axis anisotropy favors the multiple-$Q$ state, such as the skyrmion lattice, the multiple-$Q$ conical and vertical spirals. For $A$ around $-0.1$, we have the vertical spiral, skyrmion lattice and ferromagnetic state upon increasing field, which is the same as that in systems with the DM interaction in two dimensions. \cite{Bogdanov94} Since the transition between the skyrmion lattice and the ferromagnetic state is of the first order, there are skyrmion and antiskyrmion excitations in the ferromagnetic background. Our phase diagram is generally consistent with that obtained by variational calculations using the $J_1$-$J_2$ Heisenberg model on a triangular lattice in Ref. \cite{leonov_multiply_2015}, except for certain fine feature that is not accounted for in our variational space. Note that the triangular lattice of spins used in Ref. \cite{leonov_multiply_2015} favors the formation of multiple-$Q$ state because the $120^\circ$ arrangement of $\mathbf{q}_i$ is commensurate with the lattice. This may be the reason why the multiple-$Q$ vertical spiral occupies wider area in the phase diagram in Ref. \cite{leonov_multiply_2015}.

To go beyond the variational calculation, we also performed unbiased Monte Carlo simulations. The Hamiltonian Eq. \eqref{eqsd1} is difficult to simulate because of the long wavelength spin texture. Here we consider two Heisenberg models with an easy axis anisotropy on lattice:
\begin{equation}\label{eqs3abc}
\mathcal{H}=-J_{ij}\sum _{\langle i,j \rangle}\mathbf{S}_i\cdot \mathbf{S}_j-\mathbf{H}_a \sum _i \mathbf{S}_i+A\sum _i {S}_{z,i}^2,
\end{equation}
the $J_1$-$J_2$-$J_3$ model on a square lattice and $J_1$-$J_3$ model on a triangular lattice, as schematically shown in Fig. \ref{f1}. We choose the square lattice to show explicitly that the triangular lattice of spin is not necessary to stabilize the skyrmion lattice. For the square lattice case, we choose $J_2$ such that $J_1-4 J_2+16 J_3=0$, so that the spatial anisotropy vanishes to the order of $q^4$, in order to realize the Hamiltonian Eq. \eqref{eqs2}. To compare with the variational results in Fig. \ref{f6}, we normalize the magnetic field and temperature in terms of the saturation field at $A=0$, $H_s(A=0)=J(Q_0)-J(0)$.

We use the periodic boundary condition in the Monte Carlos simulations. We first anneal the system from a paramagnetic state to the target temperature by gradually reducing the temperature. The total MCS for annealing is about $10^6$. Then we thermalize the system using $5\times 10^6$ MCS and another $5\times 10^6$ MCS for measurement. The typical system size is about $72\times 72$ and $60\times 60$. We have also used $60\times 48$ in order to accommodate the triangular lattice of skyrmion. The phase diagrams obtained by Monte Carlo simulations for the $J_1$-$J_2$-$J_3$ model and the $J_1$-$J_3$ model are presented in Fig. \ref{f7}. The phase boundary is determined by analyzing the spin structure factor, spin susceptibility and specific heat as a function of magnetic field and temperature. The arrows denote the phase boundary obtained by variational calculations in Fig. \ref{f6} and they are in reasonable agreement with the Monte Carlo simulations. The discrepancy is caused by the that fact that $Q_0$ is not small because of the limitation of the Monte Carlo simulations. Several typical spin configurations and the corresponding structure factor $\langle S_x(\mathbf{q})S_x(-\mathbf{q})\rangle$ and skyrmion topological charge density, $\rho_s(\mathbf{r})$, for the vertical spiral, conical spiral, skyrmion lattice and the skyrmion liquid are displayed in Fig. \ref{f8}. In the conical spiral, there is only one dominant modulation and the modulation in other directions is weak by several orders of magnitude, because the modulations in other directions are suppressed by to the square spatial anisotropy introduced by the square lattice of spins. The square anisotropy does not favor the multiple-$Q$ conical spiral with three $\mathbf{q}_i$ vectors rotated by $120^\circ$. At high $T$, all the ordered phases are destroyed by thermal fluctuations and we have field-induced ferromagnetic state. Near the phase boundary between the skyrmion lattice and field-induced ferromagnetic state, skyrmions and antiskyrmions form liquid-like structure [Fig. \ref{f8}.]. The averaged structure factor shows a ring-like structure, indicating a liquid-like behavior of skyrmion and antiskyrmion.

\section{Dynamics}\label{Sec7}

\begin{figure}[t]
\psfig{figure=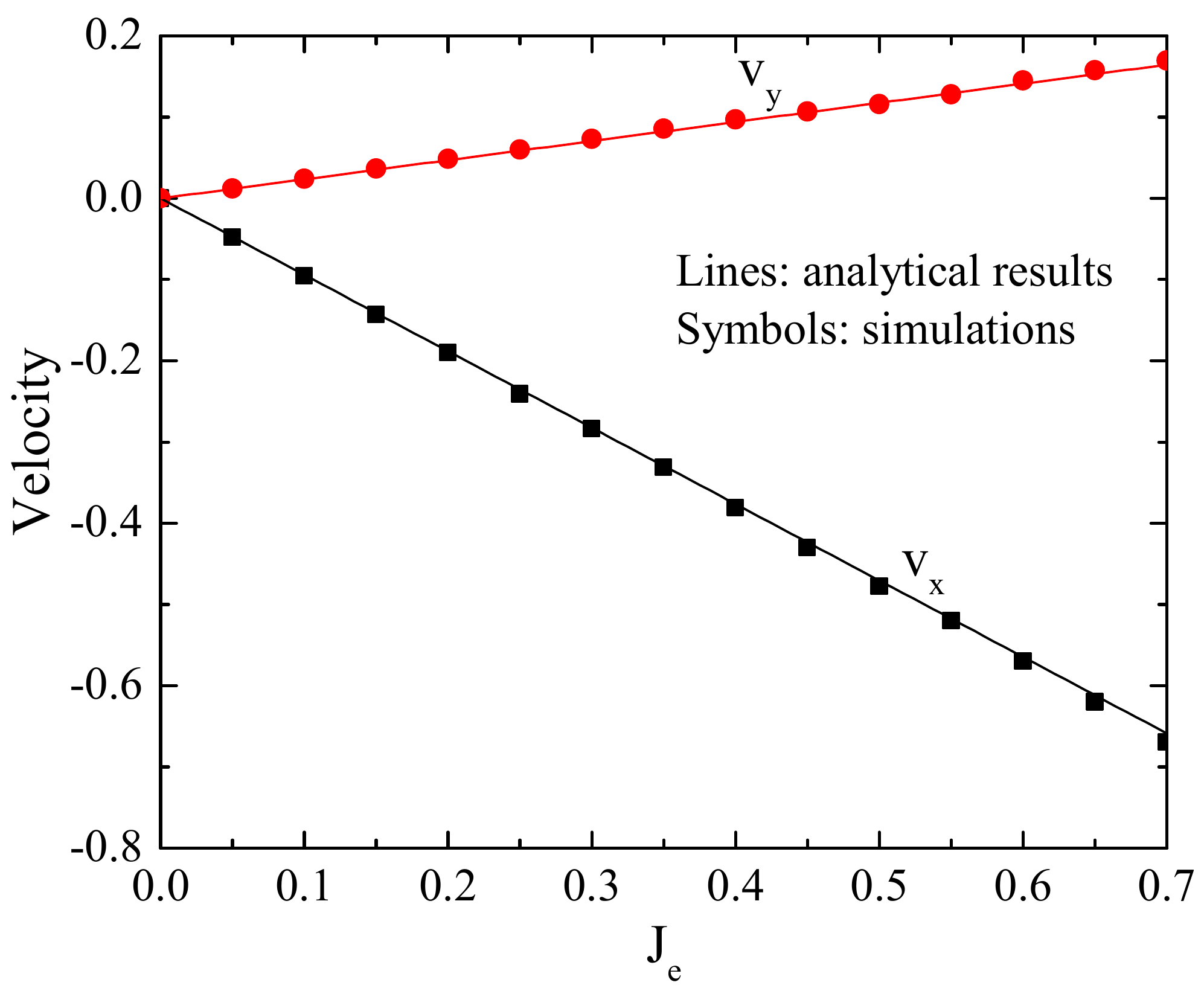,width=\columnwidth}
\caption{(color online) Skyrmion velocity as a function of current $J_e$. Lines are the results according to Eq. \eqref{eqse3} and symbols are numerical simulation of Eq. \eqref{eqseeom1}. 
} \label{f9}
\end{figure}

A skyrmion is a particle-like excitation and can be driven by external fields, such as current. The current driven dynamics of skyrmions is particularly interesting for the potential application in spintronic devices. Here we study the dynamics of a single skyrmion in the ferromagnetic state under a dc current drive. There are two Goldstone modes for a skyrmion described by Eq. \eqref{eqs2}, with one being the translational motion of the skyrmion (translational mode) and the other being the global rotation of spin along the magnetic field axis (rotation mode). Other modes associated with the deformation of skyrmion is gapped and is not accounted for if we treat skyrmion as a rigid particle. The rigidity of the skyrmion is defined based on the out-of-plane component of the spin because of the existence of the $U(1)$ symmetry associated with the helicity for the in-plane components. The translational mode can be regarded as orbital degree of freedom of a skyrmion and the rotational mode can be regarded as the spin degree of freedom of a skyrmion. In ISM with competing interactions, skyrmions have both the spin and orbital degree of freedoms, which is different from the skyrmions in chiral magnets, where skyrmions only have the orbital degree of freedom.  For a skyrmion at rest, the translational mode and the rotation mode are orthogonal to each other because they are the eigen modes of small perturbation. Generally when a skyrmion moves, these two modes hybridize and both of them are excited. There is a spin-orbit coupling for skyrmions and we will explore the interesting consequences of the spin-orbit coupling below.

\begin{figure}[b]
\psfig{figure=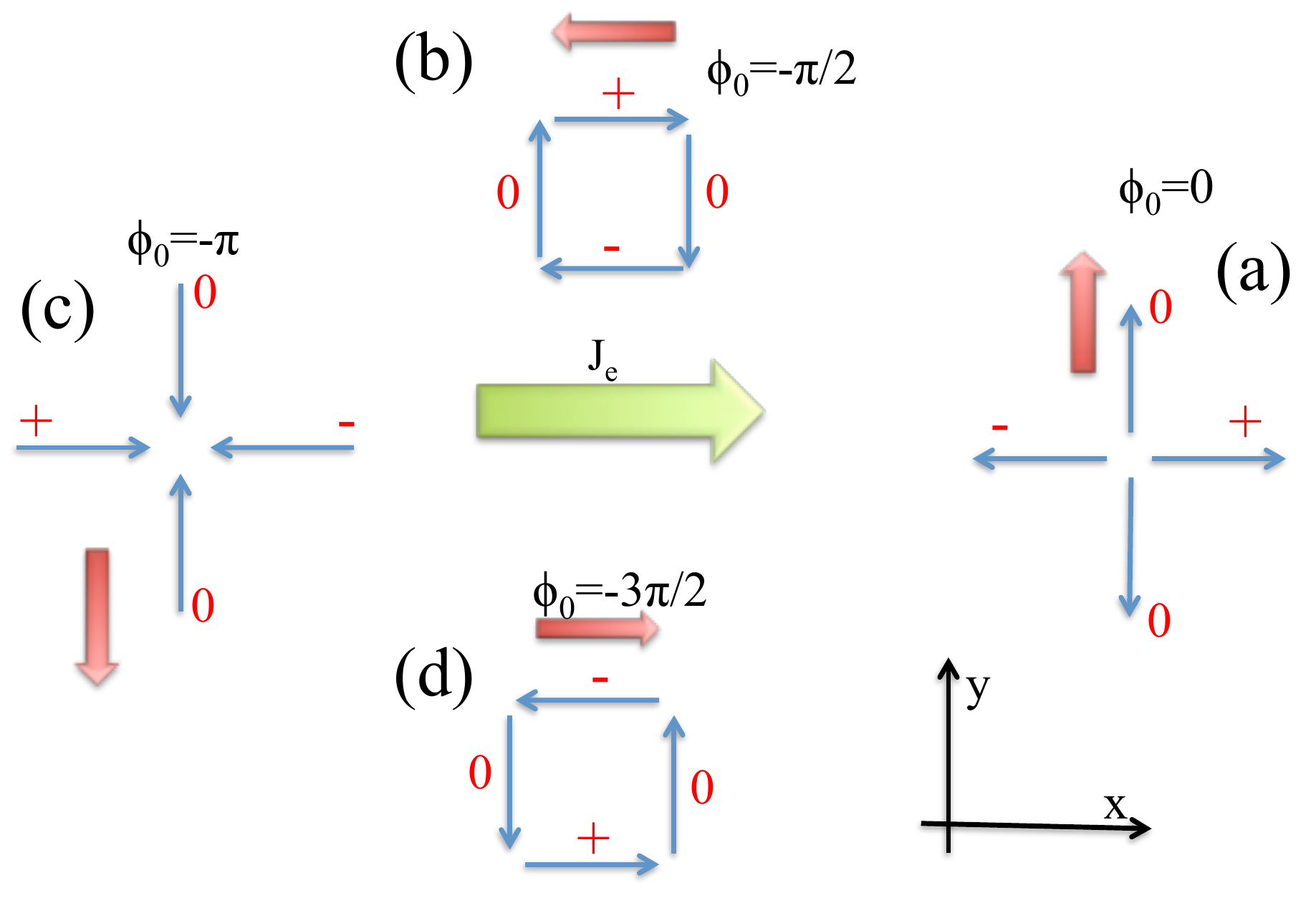,width=\columnwidth}
\caption{(color online) Schematic view of the skyrmion motion under a spin Hall torque when the rotational mode of the skyrmion is excited. The blue arrows denote the spin direction and the sign ($\pm$, $0$) close to them represent the effective magnetic field $\mathbf{H}_{\mathrm{SHE}, z}$ along the $z$ direction. The red arrows represent the direction of motion of the skyrmion.The helicity of the skyrmion is also shown. The green arrow is the external current direction and the inset is the coordinate system.  A movie based on numerical calculations is shown in Ref. \onlinecite{supplementS2S}.
} \label{f10}
\end{figure}

Let us first consider the motion of a single skyrmion in the ferromagnetic background driven by an adiabatic spin transfer torque described by the equation of motion for spins \cite{Bazaliy98,Li04,Tatara2008}
\begin{equation}\label{eqseeom1}
\partial_t\mathbf{S} =-\gamma  \mathbf{S}\times \mathbf{H}_{\mathrm{eff}}+\alpha  \mathbf{S}\times \partial_t\mathbf{S} +\frac{\hbar\gamma}{2e}(\mathbf{J}_e\cdot\nabla)\mathbf{S},
\end{equation}
where $\gamma$ is the  gyromagnetic ratio, $\mathbf{H}_{\mathrm{eff}}\equiv-\delta\mathcal{H}/\delta \mathbf{S}$ is the effective field with $\mathcal{H}$ defined in Eq. \eqref{eqs2}, $\mathbf{J}_e$ is the electric current and $\alpha$ is the Gilbert damping. We solve Eq. \eqref{eqseeom1} numerically using the method in Ref. \onlinecite{Serpico01} with the periodic boundary condition. We discretize the system into square mesh with the mesh size $\Delta_r=0.2$. We calculate the center of mass of a skyrmion defined as
\begin{equation}\label{eqseeom2}
\mathbf{R}=\frac{\int dr^2\mathbf{r}\mathbf{S}\cdot(\partial_x\mathbf{S}\times\partial_y\mathbf{S})}{\int dr^2\mathbf{S}\cdot(\partial_x\mathbf{S}\times\partial_y\mathbf{S})}.
\end{equation}
According to the calculations, for a weak current $J_e<<1$ and weak damping $\alpha<<1$ relevant for real materials, only the translational mode is excited for a spatially isotropic system. This can be seen from the term $\partial_t\mathbf{S}-\hbar\gamma (\mathbf{J}_e\cdot \nabla) \mathbf{S}/(2e)$ in Eq. \eqref{eqseeom1}, where the adiabatic spin transfer torque couples directly to the translational mode. In this case, we can use the Thiele's collective coordinate approach and use the ansatz $\mathbf{S}(\mathbf{r}-\mathbf{v} t$) with $\mathbf{v}$ the skyrmion velocity to describe the dynamics of spins. \cite{Thiele72} The equation of motion for a skyrmion is
\begin{equation}\label{eqse3}
N_s \hat{z}\times \left(\mathbf{v}+ \frac{ \hbar \gamma }{2 e}\mathbf{J}_e\right)=-\alpha  \eta  \mathbf{v}.
\end{equation}
which is the same as that in systems with DM interaction. Here $\eta =\int' \left(\partial_\mu \mathbf{n}\right)^2  dr^2/(4\pi)$ with $\mu=x,\ y$ is the form factor. In Fig. \ref{f9}, we compare the numerical results to that in Eq. \eqref{eqse3} and they agree with each other perfectly, which indicates that only the translational mode is dominant.

The rotational mode can be excited by the spin transfer torque in the presence of other coupling terms in addition to the terms in Eq. \eqref{eqs2}. Here we consider spatial anisotropy due to the underlying spin lattice. We study numerically the dynamics of a skyrmion in the $J_1$-$J_3$ Heisenberg model on a triangular lattice with $Q=2\pi/27$, where the six fold spatial anisotropy term in Eq. \eqref{eqs7} mixes the rotational mode with the translational mode. The skyrmion moves along a straight line with a velocity component perpendicular to the current. The helicity of the skyrmion changes linearly with time in addition to the translational motion, see Ref. \onlinecite{supplementS2S} for a movie. There is a spin (rotational mode)-orbit (translational mode) coupling for the skyrmion motion. The equation of motion is no longer described by Eq. \eqref{eqse3}. An analysis based on internal modes of a skyrmion will be published elsewhere.

\begin{figure}[t]
\psfig{figure=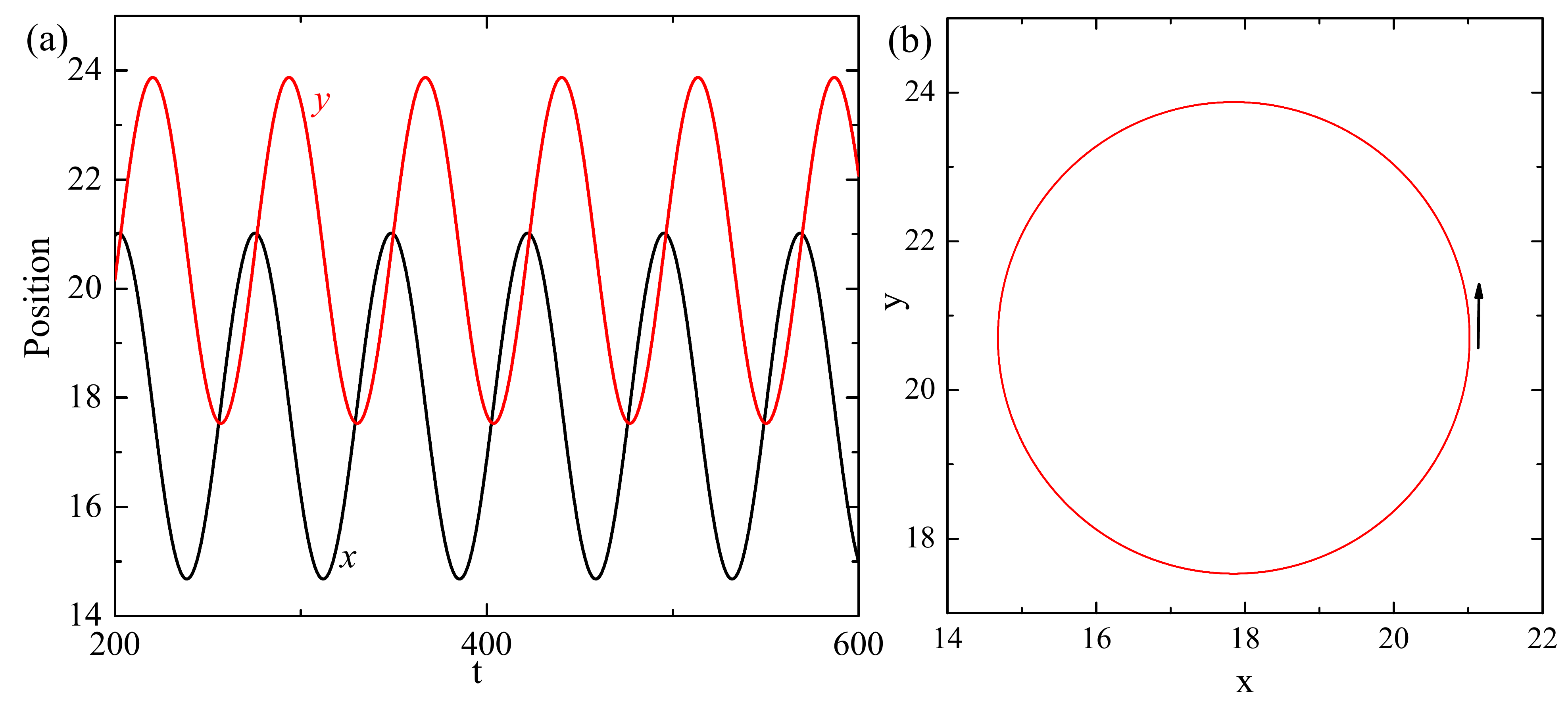,width=\columnwidth}
\caption{(color online) (a) Time dependence of the skyrmion center of mass when a skyrmion is driven by a torque due to the spin Hall effect, and (b) the corresponding trajectory. Here  $H_a=0.6$ and $\alpha=0.2$. The current is $J_e=0.2$ and is along the $x$ direction,
} \label{f11}
\end{figure}

\begin{figure*}[t]
\psfig{figure=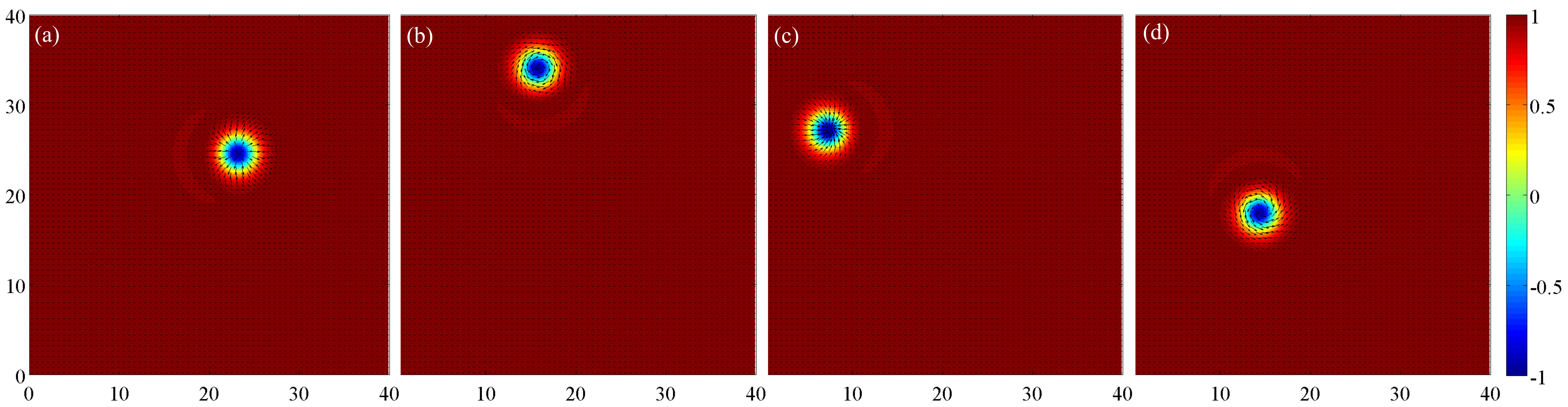,width=18cm}
\caption{(color online) Snapshots of skyrmion configuration at (a) right, (b) top, (c) left, (d) bottom of the circular trajectory. The helicity is locked with the position and changes continuously with time. Here $J_e=0.1$, $H_a=0.6$ and $\alpha=0.2$. The current is along the $x$ direction. The halos in the figures are caused by the distortion of the skyrmion due to the field gradient generated by the spin Hall torque. See Ref. \onlinecite{supplementS2S} for a movie.
} \label{f12}
\end{figure*}

Next we study the skyrmion motion driven by a torque due to the spin Hall effect, which couples directly to the rotational mode as will be shown below. The dynamics of spins is governed by \cite{emori_current-driven_2013,ryu_chiral_2013,PhysRevB.90.184427,liu_spin-torque_2012}
\begin{equation}\label{eqse4}
\partial_t\mathbf{S} =-\gamma  \mathbf{S}\times \mathbf{H}_{\mathrm{eff}}+\alpha  \mathbf{S}\times \partial_t\mathbf{S} +\frac{\hbar \gamma \theta_{\mathrm{sh}}}{2e d}\mathbf{S}\times \left[\mathbf{S}\times \left(\hat{z}\times\mathbf{J}_e\right)\right],
\end{equation}
where $\theta_{\mathrm{sh}}$ is the spin Hall angle and $d$ is the film thickness. The torque due to the spin Hall effect can be described by an effective magnetic field $\mathbf{H}_{\mathrm{SHE}}\propto \mathbf{S}\times \left(\hat{z}\times\mathbf{J}_e\right)$. We assume that the current is along the $x$ direction. According to the simulations, the rotational mode is excited with the helicity changing linearly in time, $\phi_0=\omega t$. For a skyrmion with helicity $\phi_0=0$ shown in Fig. \ref{f10}, there is magnetic field gradient of ${H}_{\mathrm{SHE}, z}$ along the $x$ direction, which generates a force along the same direction. Because of the Magnus force, the skyrmion moves perpendicularly to the force and it moves along the $y$ direction. \cite{szlin13skyrmion2,Iwasaki2013} When the helicity becomes $-\pi/2$, the skyrmion experiences a field gradient ${H}_{\mathrm{SHE}, z}$ in the $y$ direction and the skyrmion moves in the negative $x$ direction. Similarly when $\phi_0=-\pi$, skyrmion moves along the negative $y$ direction and for $\phi_0=-3\pi/2$, it moves along the positive $x$ direction. Therefore the skyrmion moves along a circle in the presence of a spin Hall torque generated by a dc current, and its helicity is locked to the position.

To verify the above picture, we derive the equation of motion for a rigid skyrmion with a fixed helicity using the Thiele's approach \cite{Thiele72,szlin2015Spiral2Skyrmion}
\begin{equation}\label{eqse5}
N_s \hat{z}\times \mathbf{v}+\alpha  \eta  \mathbf{v}=\frac{\hbar \gamma \theta_{\mathrm{sh}}}{2e d} J_e \mathbf{Y},
\end{equation}
with
\begin{equation}
Y_\mu=\left(\hat{z}\times \hat{\mathbf{J}} \right)\cdot \int'\left(\partial_\mu \mathbf{n}\times\mathbf{n} \right)d r^2/(4\pi),
\end{equation}
and $\hat{\mathbf{J}} $ a unit vector along the current direction. Without loss of generality, we assume current is along the $x$ direction, $\mathbf{J}_e=J_e\hat{x}$. Then
\begin{equation}
\mathbf{Y}=Y_0\hat{l}_{\pi-\phi_0},
\end{equation}  
\begin{equation}
Y_0=\frac{\pi}{4}+\frac{1}{8}\int_0^{\infty} \sin(2\theta)d r,
\end{equation}
where $\hat{l}_{\pi-\phi_0}$ is a unit vector with a polar angle $\pi-\phi_0$. The equation of motion depends explicitly on the helicity. In the other words, the spin Hall torque couples to the rotational mode. The velocity perpendicular (parallel) to $\mathbf{Y}$, $v_\perp$ ($v_\parallel$) is
\begin{equation}
v_\perp=-N_s Y_0 \hbar \gamma \theta_{\mathrm{sh}} J_e/[2ed(N_s^2+(\alpha\eta)^2)],
\end{equation}
\begin{equation}
v_\parallel=\alpha \eta Y_0 \hbar \gamma \theta_{\mathrm{sh}} J_e/[2ed(N_s^2+(\alpha\eta)^2)].
\end{equation}
For a weak damping $\alpha\ll 1$, the velocity is almost perpendicular to $\hat{l}_{\pi-\phi_0}$. When the rotational mode is excited, $\phi_0=\omega t$, the skyrmion move along a circle.

We solve numerically Eq. \eqref{eqse4} with a skyrmion in the ferromagnetic background as an initial state and find the skyrmion performs a circular motion with continuously changing its helicity as shown in Figs. \ref{f11} and \ref{f12}, consistent with the above analysis. To determine the $\omega$, one needs to the study the hybridization between the translational mode and rotational mode.

\begin{table*}
\caption{\label{tbl1} Comparison between skyrmions in inversion-symmetric magnets and chiral magnets.}
\begin{ruledtabular}
\begin{tabular}{|c|c|c|}
{\bf{Properties of skyrmions}} & {\bf{Inversion-symmetric magnets}} & {\bf{Chiral magnets}} \\ \hline
Energy of skyrmion and antiskyrmion &  Degenerate & Non-degenerate \\ \hline
Skyrmion size in two dimenions &  Diverges at the saturation field & finite \\ \hline
Helicity &  Arbitrary & Fixed by the DM vector \\ \hline
Pairwise interaction as a function of separation &  Nonmonotonic and depends on helicity & Monotonic\\ \hline
Goldstone modes & \pbox{20cm}{Translation in space and \\ global rotation of spin along the magnetic field axis. \\ There can be coupling between these two modes.}  & Translation in space \\ \hline
Equation of motion driven by a dc spin Hall torque &  Move along a circle & Move along a straight line\\
\end{tabular}
\end{ruledtabular}
\end{table*}

\section{Discussions}\label{Sec8}

The derivation of the effective theory Eq. \eqref{eqs2} are based on the expansion of ordering wave vector around zero valid when $Q a<<1$. For a large $Q$, such as expansion is not allowed. However we can expand the $Q$ around a $Q_0$ that is commensurate with the spin lattice. For example, in the frustrated Heisenberg model with nearest neighbor antiferromagnetic interaction on triangular lattice, the ground state is a spiral with $Q_0=2\pi/3$. In the present of high order weak interactions, such as interlayer FM coupling \cite{PhysRevB.81.104411}, the ordering $Q$ deviates slightly from $Q_0=2\pi/3$, $Q=Q_0+\delta$ with $\delta\ll 1$. We need to introduce three order parameters $\mathbf{S}_i$, $i=1, 2, 3$, defined on the three sublattices and do the same expansion in $\delta$. The resulting Ginzburg-Landau energy contain three spin vector field $\mathbf{S}_i(\mathbf{r})$ with coupling among them.  

We then compare the skyrmions in inversion-symmetric magnets to that in chiral magnets and highlight the differences. For magnet with inversion symmetry, the skyrmions with opposite winding direction or skyrmion and antiskyrmion have the same energy. Close to the phase boundary between the skyrmion lattice and ferromagnet, there is skyrmion-antiskyrmion liquid like phase. While for the chiral magnets, the winding direction is determined by the sign of the DM vector, so either skyrmions or antiskyrmions are stabilized. In ISM, there is $U(1)$ symmetry associated with the global rotation of spin along the magnetic field direction. Therefore the helicity of skyrmion is not determined. In chiral magnets, the DM interaction removes the $U(1)$ symmetry and the helicity is selected by the direction of the DM vector. For the DM interaction generated by the Dresselhaus spin orbit interaction, a Bloch skyrmion is stabilized with helicity $\phi_0=\pi/2$, while for the DM interaction generated by the Rashba spin orbit interaction, a N\'{e}el skyrmion is stabilized with helicity $\phi_0=0$. For a single skyrmion in ISM with competing interactions, the canting angle of spin with respect to field direction decreases from $\pi$ (antiparallel to the field) to zero (parallel to field) and then oscillates around zero indicating a reversal of helicity. This also implies that the pairwise interaction between skyrmions are nonmonotonic as a function of distance between skyrmions and depends on the helicity. Meanwhile the decay length increases when the magnetic field approach the saturation field from above meaning the skyrmion size diverges. In the case of chiral magnets, the canting angle decreases monotonically and the skyrmion size is always finite. The pairwise interaction between skyrmions is monotonic. 

For skyrmions in ISM, we have two Goldstone modes with one being the translational motion of skyrmion and the other rotation of spin along the $z$ axis due to the $U(1)$ symmetry. These two Goldstone modes can be hybridized by spatial anisotropy introduced by the spin lattice or by torque due to the spin Hall effect. Thus the skyrmions in ISM gain a ``spin" degree of freedom and the spin of a skyrmion can couple to its orbital degree of freedom. When both the rotational and translational modes are excited, the equation of motion for skyrmion is no longer described by the original Thiele's equation. Moreover in the presence of a torque due to the spin Hall effect induced by a dc current, we found that the skyrmion moves along a circle. While for the chiral magnets, we only have the translational mode as the Goldstone mode and the equation of motion of skyrmion can be described by the Thiele's equation. Table \ref{tbl1} summarizes the differences between skyrmions in ISM and chiral magnets.

A measurable quantity associated with the helicity $\phi_0$ is the toroidal moment $\mathcal{T}(\phi_0)\equiv \int dr^3 \mathbf{r}\times\mathbf{S}\propto\sin\phi_0$ \cite{Spaldin2008}. The toroidal moment is allowed because the spatial inversion symmetry and time reversal symmetry are broken in the presence of skyrmions, although the original Hamiltonian Eq. \eqref{eqs2} has these two symmetries. The toroidal moment density oscillates when one moves away from the skyrmion center. The toroidal moment couples with electric current, which points to a possible way to tune the helicity of the skyrmion lattice. When the rotational mode of skyrmion is excited, $\mathcal{T}$ oscillates with time which can be measured experimentally.

Competing interactions and a weak easy axis anisotropy in magnets are sufficient to stabilize skyrmion lattice in the universal long wavelength limit. In this universal region, the skyrmion lattice does not depend on the underlying spin lattice and the microscopic origin of the competing interactions. This suggests that the skyrmion lattice may be a ubiquitous state in magnetic materials and also provides a guidance for experimental search of new skyrmion-hosting materials. Several possible candidates such as, $\mathrm{NiGa_2S_4}$, $\alpha$-$\mathrm{NaFeO_2}$, $\mathrm{Fe_xNi_{1-x}Br_2}$, have already been proposed in Refs. \onlinecite{PhysRevLett.108.017206, leonov_multiply_2015}. The easy axis anisotropy generally favors multiple-$Q$ ordering, such as multiple-$Q$ vertical spiral and conical spiral, suggesting that the multiple-$Q$ ordered states may be ubiquitous in magnets with competing interactions.

To summarize, we derive a Ginzburg-Landau theory  to describe the skyrmion physics in inversion-symmetric magnets with competing interactions. The general theory is valid for any classical Heisenberg model with competing interactions in the long wavelength limit. Our theory shows that the stabilization of skyrmion lattice does not depend on the symmetry of the underlying spin lattice and an easy axis anisotropy is sufficient to stabilize a skyrmion lattice. We also demonstrate the unusual properties of skyrmions, such as structure of a single skyrmion, nonmonotonic interaction between skyrmions and dynamics, and compare to those in chiral magnets with Dzyaloshinskii-Moriya interaction. Remarkably, the skyrmions in inversion-symmetric magnets gain a  ``spin" degree of freedom in addition to the usual orbital degree of freedom. The spin-orbit coupling in the skyrmion motion gives rise to highly nontrivial results that are not shared by skyrmions in chiral magnets. Our theory will trigger further interests in skyrmions in inversion-symmetric magnets and in searching for skyrmions in these materials.

\begin{acknowledgments}
The authors are indebted to Mohit Randeria and Cristian D. Batista for helpful discussions. Computer resources for numerical calculations were supported by the Institutional Computing Program at LANL. This work was carried out under the auspices of the NNSA of the US DOE at LANL under Contract No. DE-AC52-06NA25396, and was supported by the US Department of Energy, Office of Basic Energy Sciences, Division of Materials Sciences and Engineering. 
\end{acknowledgments}

\bibliography{reference}

\end{document}